\documentclass[12pt]{article}
\usepackage{natbib}
\usepackage{bm}
\usepackage{amsmath}
\usepackage{amssymb}
\usepackage{dsfont}
\usepackage{tikz}
\usepackage{soul}
\usepackage{url}
\usepackage{longtable}
\usepackage{float}
\usepackage{subfigure}
\usepackage{enumerate}
\usetikzlibrary{intersections,backgrounds,arrows,calc,shapes,angles,positioning}

\usepackage[all]{xy}  

\usepackage{chngcntr}
\counterwithin{figure}{section}    
\counterwithin{table}{section}

\newcommand{\blind}{1}

\usepackage{verbatim}
\usepackage[all]{xy}
\xyoption{arc}
\usepackage{xkeyval} 
\usepackage{tikz}
\usepackage{etoolbox}
\usepackage{multirow}
\usepackage{caption}
\usetikzlibrary{intersections}

\date{}
\begin{document}
\def\spacingset#1{\renewcommand{\baselinestretch}%
{#1}\small\normalsize} \spacingset{1}

\if1\blind
{
  \title{\bf Bayesian Integrative Analysis and Prediction with Application to Atherosclerosis Cardiovascular Disease}
  \author{Thierry Chekouo
  \thanks{corresponding author}\hspace{.2cm}\\
    \small{Department of Mathematics and Statistics, University of Calgary, AB}\\
    \small{and} \\
    Sandra E. Safo \thanks{Authors contributed equally}
    \\
    \small{Division of Biostatistics, University of Minnesota, Minneapolis, MN}
    }
  \maketitle
} \fi

\if0\blind
{
  \bigskip
  \bigskip
  \bigskip
  \begin{center}
    {\LARGE\bf Bayesian Integrative Analysis and Prediction with Application to Atherosclerosis Cardiovascular Disease}
\end{center}
  \medskip
} \fi

\bigskip

\begin{abstract}
Cardiovascular diseases (CVD), including atherosclerosis CVD (ASCVD), are multifactorial diseases that present a major economic and social burden worldwide. Tremendous efforts have been made to understand traditional risk factors for ASCVD, but these risk factors account for only about half of all cases of ASCVD.  It remains a critical need to identify nontraditional risk factors (e.g., genetic variants, genes) contributing to ASCVD. Further, incorporating functional knowledge in prediction models have the potential to reveal pathways associated with disease risk. We propose Bayesian hierarchical factor analysis models that associate multiple omics data, predict a clinical outcome, allow for prior functional information, and can accommodate clinical covariates.  The models, motivated by available data and the need for other risk factors of ASCVD, are used for the integrative analysis of clinical, demographic, and multi-omics data to identify genetic variants, genes, and gene pathways potentially contributing to 10-year ASCVD risk in healthy adults. Our findings revealed several genetic variants, genes and gene pathways that were highly associated with ASCVD risk. Interestingly, some of these have been implicated in CVD risk. The others could be explored for their potential roles in CVD. Our findings underscore the merit in joint association and prediction models.   \\
  
\end{abstract}

\noindent%
	\vfill
\newpage
\spacingset{1.5} 
	\section{Introduction}
\noindent Cardiovascular  diseases (CVD) including atherosclerosis CVD (ASCVD) are multifactorial, and are caused by a combination of many factors such as genetic, biological, and environmental factors. 
Many research applications and methods continue to demonstrate that a better  understanding of the pathobiology of CVD and other complex diseases necessitates analysis techniques  that go beyond the use of established traditional factors to one that integrate  clinical and demographic data with molecular and/or functional data. For instance, in a recent CVD research, the authors in \citep{cvdriskproteome}  integrated  genomics data with CVD-risk associated proteins in human plasma and they identified many genetic variants (not previously described in the literature) which, on average, accounted for 3 times the amount of variation in CVD explained by common clinical factors, such as age, sex, and diabetes mellitus \citep{cvdriskproteome}. Given that  
environmental risk factors for ASCVD (e.g., age, gender, hypertension) account for only half of all cases of ASCVD \citep{Bartels:2012}, it remains a critical need to identify other biomarkers contributing to ASCVD. It is also recognized that leveraging existing sources of biological information including gene pathways can guide the selection of clinically meaningful biomarkers, and reveal pathways associated with disease risk.  In this study, we build models for  associating multiple omics data and simultaneously predicting a clinical outcome while considering prior functional information. The models are motivated by, and are used for, the integrative analysis of multi-omics and clinical data from the Emory Predictive Health Institute, where the goal is to identify genetic variants, genes, and gene pathways potentially contributing to 10-year ASCVD risk in healthy patients. 

Our approach builds and extends recent research on joint integrative analysis and prediction/classification methods \citep{kan2015multi,Luo2014,li2018integrative, SIDA:2019}. Unlike two step approaches-- integrative analysis followed by prediction or classification-- these joint methods link the problem of assessing associations between multiple omics data to the problem of predicting or classifying a clinical outcome. The goal is then to identify  linear combinations of variables from each omics data that are associated and are also predictive of a clinical outcome. For instance, the authors in \citep{Luo2014,li2018integrative,  SIDA:2019} combined canonical correlation analysis with classification or regression methods for the purpose of  predicting a binary  or continuous response. 
\cite{Luo2014}, used  a regression formulation of canonical correlation analysis (CCA) and proposed a joint method for obtaining canonical correlation vectors and predicting an outcome from the canonical correlation vectors. 
The authors in \citep{SIDA:2019} proposed a joint association and classification method that combines CCA and linear discriminant analysis, and uses the normalized Laplacian of a graph to smooth the rows of the discriminant vectors from each omics data. This approach encourages predictors that are connected and behave similarly to be selected or neglected together.  These joint association and prediction methods have largely been developed from a frequentist perspective. While there exist Bayesian factor analysis or CCA methods to integrate multi-omics data types \citep{Mo2018,Klami:2013}, we are not aware of any  Bayesian method that has been developed for joint association and prediction of multiple data types. \cite{Mo2018} extended icluster into a Bayesian framework, and  \cite{Klami:2013} developed a Bayesian CCA model by imposing a group-wise sparsity for inter-battery factor analysis (IBFA).

In the present work, we extend current frequentist approach for integrative analysis and prediction methods to the Bayesian setup. Our contributions are multi-fold. First,  we adopt a factor analysis (FA) approach to integrate multiple data types as done in \citep{ Shen2009,Ronglai2013,Klami:2013}. This allows us to reduce simultaneously the dimension of each data type to a shared component  with a much reduced number of features (or components).  On the other hand, our formulation of the problem is different from existing methods as it is defined  in a Bayesian hierarchical setting which has the flexibility to incorporate other functional information through prior distributions. Our approach uses ideas from Bayesian sparse group selection to identify simultaneously active components and important features within components using two nested layers of binary latent indicators. Second, our approach is more general; we are able to account for the active shared components as well as the  individual latent components for each data type.  Third, we incorporate in a unified procedure, clinical responses that are associated with the shared components, allowing us to  evaluate the predictive performance of the shared components.  Moreover, our formulation makes it easy to include other covariates without enforcing sparsity on their corresponding coefficients.  Including other available covariates may inform the shared components, which in turn may result in better predictive performance of the shared components. Fourth, unlike most integrative methods that use FA, our prior distributions can be extended to incorporate external grouping information; this allows  us to determine most important groups of features that are associated with important components.  Incorporating prior knowledge
about grouping information has the potential of identifying functionally meaningful
variables (or network of variables) within each data type for improved prediction performance of the response variable. 
In our estimation, we implement a Markov Chain Monte Carlo (MCMC) algorithm for posterior inference that employs a  partially collapsed Gibbs sampling \citep{David2008} to sample the latent variables and the loadings.

	\subsection{Motivating Application}
There is an increased interest in identifying ``nontraditional" risk factors (e.g., genes, genetic variants) that can predict CVD and ASCVD. This is partly so because CVD has become one of the costliest chronic disease, and continue to be the leading cause of death in the U.S. \citep{CVD:2016}.  It is projected that nearly half of the U.S. population will have some form of cardiovascular disease by 2035 and will cost the economy about $\$2$ billion/day in medical costs \citep{CVD:2016}. There have been significant efforts in understanding the risk factors associated with ASCVD, but research suggests that the 
environmental risk factors for ASCVD (e.g., age, gender, hypertension) account for only half of all cases of ASCVD \citep{Bartels:2012}.  Finding other risk factors of ASCVD and CVD unexplained by traditional risk factors is important and potentially can serve as targets for treatment.  

We integrate gene expression, genomics, and/or clinical data from the Emory University and Georgia Tech Predictive Health Institute (PHI) study to  identify potential biomarkers contributing to 10-year ASCVD risk among healthy patients. The PHI study, which began in 2005, is a longitudinal study of healthy employees of Emory University and Georgia Tech aimed at collecting health factors that could be used to recognize, maintain, and optimize health rather than to treat  disease. 
The goal of the PHI, the need to identify other biomarkers of atherosclerosis cardiovascular diseases, and the data available from PHI (multi-omics, clinical, demographic data), motivate our development of Bayesian methods for associating multiple omics data and simultaneously predicting a clinical outcome while accommodating prior functional information and clinical covariates.

The rest of the paper is organized as follows. In Section \ref{modelform}, we present the formulation of the model approach. In Section \ref{postinfe}, we  describe our MCMC algorithm, present how to evaluate our prediction performance, and we set hyperparameters of prior distributions. In Section \ref{simul}, we evaluate and compare our methodology using simulated data. Section \ref{appli} shows the application to atherosclerosis disease using both mRNA expression and SNP information. Finally, we conclude our manuscript in Section \ref{conclu}.
	\section{Model formulation\label{modelform}}
Our primary goal is to define an integrative approach that combines multiple data types and incorporates clinical covariates, response variables, and external grouping information. In Section \ref{FAmodel}, we introduce the Factor analysis model that relates data types. We then adopt a Bayesian variable selection technique in Section \ref{BVS} to select active components and their corresponding important features. We show how we incorporate grouping information in Section \ref{groupinfo}.  
	\subsection{Factor analysis model for integrating multiple data types and clinical outcome\label{FAmodel}}
We assume that $M$ (-omics) data types are available: $\bm{X^{(m)}}=(x_{ij}^{(m)})_{n\times p_m}$, where  $x_{ij}^{(m)}$ is the measure of subject $i$ on the feature $j$ for data type $m=1,...,M$. A factor analysis model for the $M$ data types can be written as   
 	\begin{eqnarray}\label{model1}
 	\bm{X^{(m)}}=\bm{U}\bm{A^{(m)}}+\bm{E^{(m)}}, \textrm{ for } m=1,...,M\\  	
 	\bm{U}\sim\bm{N}(\bm{0},\bm{I}_{nr}) \text{ and } \bm{E^{(m)}}\sim \mathcal{N}(\bm{0},\Psi^{(m)})	\nonumber
 	\end{eqnarray}
 	Here, $\bm{I}_{nr}$ is the identity matrix of size $nr$. $\bm{A^{(m)}}$  is a $r\times p_m$ coefficient matrix for data type $m$ with each row representing  the factor loadings for components $l=1,...,r$.  $\bm{U}$ is a $n\times r$ matrix of latent variables that connects the $M$ sets of models,  thus inducing dependencies among the data types. The latent variable is intended to explain the correlations across the data types and $\bm{E^{(m)}}$ is  the error that explains the remaining variability  unique to each data type. The integrated data matrix 
 	$ (\bm{X}^{(1)},...,\bm{X}^{(M)})$ is then multivariate normal with mean zero and covariance matrix $\bm{\Sigma} = \bm{A}^T\bm{A}+ \bm{\Psi}$, where $\bm{A}= ( \bm{A}^{(1)},...,\bm{A}^{(M)})$ and $\bm{\Psi}$ is the diagonal block matrix of $\Psi^{(m)}$, $m=1,...,M$. As it is commonly done in factor analysis, we assume $\Psi^{(m)}=Diag(\sigma_1^{2(m)},\sigma_2^{2(m)},....,\sigma_{p_m}^{2(m)})\otimes I_n$ where $\sigma_j^{2(m)}$ is the variance of  feature $j$  in data set $m$.   Model (\ref{model1}) is also called the  \textit{Gaussian latent variable model}. This model has been extensively used for integrative clustering of multiple genomic datasets \citep{ Shen2009,Ronglai2013,  Prabhakar2014}
 	
 In this paper, we are  interested in examining the association between the two sets of features and performing predictive modeling of the response in an integrative way.   Hence, in addition to model (\ref{model1}), we treat the clinical outcome $\bm{y}_{n\times 1}$ as another variable set, and is related to feature matrices $ \bm{X}^{(1)},...,\bm{X}^{(M)}$  through the latent variables $\bm{U}$. Thus, the clinical equation can be written as $\bm{y}=\bm{X}^{(0)}=\alpha_0+\bm{U}A^{(0)}+\bm{E^{(0)}}$, where  $A^{(0)}$ is a coefficient vector of the effect component $l$, $l=1,\ldots,r$, has  on the outcome; $\alpha_0$ is the intercept and   $\bm{E^{(0)}}\sim \mathcal{N} (0,\sigma^{2(0)}I_n)$ is the error term. We also assume that there is another data type $m=M+1$ which corresponds to the  set of clinical covariates.

 		\subsection{Bayesian variable selection using latent binary variables\label{BVS}}
 		
 		To determine active components for each data type (including the outcome), we define an $r$-binary latent variable vector, $\bm{\gamma^{(m)}}=(\gamma^{(m)}_{1},,...,\gamma^{(m)}_{r})$, given by  $\gamma^{(m)}_{l}=1$ if  component $l=1,...,r$ is active in data type $m$ and 0 otherwise for $m=0,1,...,M+1$. We assume that for $m=M+1$, $\gamma^{(M+1)}_{l}=1$ for every $l$  so that all components are potentially active for clinical covariates.  
 		
 		We are not only interested in selecting the active components but we also want to identify variables that significantly contribute to active components. Unlike the single layer of indicators used in the group-wise sparsity for inter-battery factor analysis (IBFA) \citep{Klami:2013}, we use two nested layers of binary indicators to denote whether a component or a variable is selected or not. A similar idea was recently used  for Bayesian sparse group selection in regression models \citep{Chen2016}.  At the component level, for every $m=1,...,M$,  $\bm{\gamma^{(m)}}$ indicates which components are active for data type $m$.  At the variable level, $\bm{\eta_{j}}^{(m)}=(\eta_{1j}^{(m)},...,\eta_{rj}^{(m)})$ is an indicator vector for variables in the $l$th component. Specifically, $\eta_{lj}^{(m)}=1$ if feature $j$ is selected in component $l$ in data type $m$. 
 		The prior distribution for the component indicator, $\gamma^{(m)}_{l}$, is the Bernoulli distribution with parameter $q_m$.  In the $l$th component, the prior distribution of $\eta_{lj}^{(m)}$ is dependent on the indicator $\gamma^{(m)}_{l}$ and is defined as
 		\begin{eqnarray}
 	\eta^{(m)}_{lj}|\gamma^{(m)}_{l} &\sim& (1-\gamma^{(m)}_{l})\delta_{0}+\gamma^{(m)}_{l}\text{Bernouilli}(q_{\eta}),
 		\end{eqnarray}
 		where $\text{Bernouilli}(q_{\eta})$ is the Bernoulli distribution with probability $q_{\eta}$. Thus, if the $l$th component is not active, then $\eta^{(m)}_{lj}=0$ for all $j$. We note that $\eta^{(m)}_{lj}=1$ for every $l$ and $m=0,M+1$ i.e the response variable and clinical covariates will be always included in the model.  
 		    To infer a sparse matrix $\bm{A}^{m}$, that is whether the $a^{(m)}_{lj}$ entry in $\bm{A}^{m}$ is important or not, we impose the following sparsity-inducing prior on the matrix  elements given the indicators $\eta^{(m)}_{lj}$ and $\gamma^{(m)}_{l}$
	\begin{eqnarray}
a^{(m)}_{lj}|\gamma^{(m)}_{l},\eta^{(m)}_{lj}&\sim&(1-\gamma^{(m)}_{l})\delta_{0}+\gamma^{(m)}_{l}[(1-\eta^{(m)}_{lj})\delta_{0}+\eta^{(m)}_{lj}\mathcal{N}(0,(\tau_{lj}^{(m)})^2\sigma_j^{2(m)})]\\\nonumber
&\sim& (1-\gamma^{(m)}_{l}\eta^{(m)}_{lj})\delta_{0}+\gamma^{(m)}_{l}\eta^{(m)}_{lj}\mathcal{N}(0,(\tau_{lj}^{(m)})^2\sigma_j^{2(m)}).
	\end{eqnarray}
	Thus, if the $j$th variable in the active $l$th component is selected, then the prior of $a^{(m)}_{lj}$ is a normal
distribution with zero mean and variance $(\tau_{lj}^{(m)})^2\sigma_j^{2(m)}$. Otherwise the coefficient $a^{(m)}_{lj}$ is  equal to 0. When $m=0$, the feature  $\bm{X}^{(0)}$ is now the outcome and $\eta_{l1}^{(0)}=\gamma^{(0)}_{l}$.  The prior distribution of the residual variance $\sigma_j^{2(m)}$ is an inverse Gamma distribution, $\sigma_j^{2(m)}\sim
IG(a_0, b_0)$. Finally, we assume that $\gamma^{(m)}_{l}\sim \text{Bernouilli}(q^{(m)})$ and $q^{(m)}$ follows a Beta distribution $Beta(a,b)$.
 	
 Our method encompasses three different scenarios: i) each component is  shared across all  data types (including the response) i.e. $\gamma^{(m)}_{l}=1$ for every $m=0,1,2,\ldots,M,M+1$ and $l=1,...,r$.  This scenario only accounts for the joint variation between the data types, and  is assumed  in standard CCA methods for high dimensional data integration such as \citep{WT:2009,Luo2014,Safo2018} ii) none of the $r$ components is shared across data types i.e. $\displaystyle \bigcap^{M}_{m=0} \{l=1,...,r;\gamma^{(m)}_{l}=1\}=\varnothing$. This scenario would be similar to the standard factor analysis for each data set independently, and accounts only for the individual variations explained for each data set. iii) some components are only shared across the data sets. Hence, our method can capture the amount of joint structure and individual structure of the data sets.   Our method then encompasses the Bayesian CCA and the JIVE (Joint and Individual variation explained) methods introduced respectively by \cite{Klami:2013} and \cite{Lock2013}. For instance, assume we have $2$ other data types (besides response and clinical covariates), $r=4$ components,  the first two components are shared (i.e $\gamma^{(m)}_{1}= \gamma^{(m)}_{2}=1$, for every $m$) and, components 3 and 4 are only associated respectively with data types 1 and 2 (i.e. $\gamma^{(1)}_{3}=\gamma^{(2)}_{4}=1$ and  $\gamma^{(2)}_{3}=\gamma^{(1)}_{4}=0$). The models for data types 1 and 2 can  then be written as $\bm{X}^{(1)}=\bm{U}\bm{A}^{(1)}+\bm{E}^{(1)}=\bm{U}_{1,2}\bm{A}_{1,2}^{(1)}+\bm{U}_{3}\bm{A}_{3}^{(1)}+\bm{E}^{(1)}$ and $\bm{X}^{(2)}=\bm{U}\bm{A}^{(2)}+\bm{E}^{(2)}=\bm{U}_{1,2}\bm{A}_{1,2}^{(2)}+\bm{U}_{4}\bm{A}_{4}^{(2)}+\bm{E}^{(2)}$, where $\bm{U}_{1,2}=(\bm{U}_{1},\bm{U}_{2})$ (resp. $\bm{A}_{1,2}$) is the matrix of the first two components of $\bm{U}=(\bm{U}_{1},\bm{U}_{2},\bm{U}_{3},\bm{U}_{4})$ (resp. $\bm{A}$). In this example, $\bm{U}_{1,2}$ is the shared component, and $\bm{U}_{3}$ and $\bm{U}_{4}$ can be considered as the individual latent structures for data types 2 and 3 respectively. 
 		
 		\subsection{Incorporating grouping information through prior distributions\label{groupinfo}}
In this section, we present how we incorporate  known group structures of the $M$ data types through prior distributions. For that, we introduce 	a design/loading matrix $\bm{P}^{(m)}=(\kappa^{(m)}_{jk})_{ (p_m\times K_m )}$ consisting of $K_m$ columns of dummy variables coding for group membership, for every $m=1,...,M$. More specifically, $\kappa^{(m)}_{jk}=1$ if marker $j$ from data type $m$ belongs to group $k=1,...,K_m$, and 0 otherwise. Here we assume that this matrix can be retrieved from external scientific knowledge. For instance, genes  can be grouped using KEGG pathway information  or other pathway/network information \citep{Qiu2013}. Similarly,  SNPs can be grouped using their corresponding genes. 

In order to incorporate the grouping information $\bm{P}^{(m)}$, we assume an additional layer from the prior of the factor loadings $\bm{A^{(m)}}$. More formally,  we incorporate the grouping information through the prior distributions of $(\tau_{lj}^{(m)})^2$ as follows
\begin{eqnarray}
(\tau_{lj}^{(m)})^2|\lambda^{(m)}_{lj}&\sim&\text{Exp}((\lambda^{(m)}_{lj})^2)\\
(\lambda^{(m)}_{lj})^2|\bm{b}^{(m)}_{l\cdot}&\sim&\Gamma(\alpha,b^{(m)}_{l0}+P^{(m)T}_j\bm{b}^{(m)}_{l\cdot}),\label{modelgrp}
\end{eqnarray}
where $P^{(m)T}_j$ denotes the $j$-th row of the matrix $\bm{P}^{(m)}$ and $\Gamma(\alpha,\beta)$ denotes the gamma distribution with shape $\alpha$ and rate $\beta$. The vector $\bm{b}^{(m)}_{l\cdot}=(b^{(m)}_{lk})_{K_m}$ represents the coefficient effects of the groups. The intercept $b_{l0}$ can be regarded as a global shrinkage hyper-parameter determining the baseline level of shrinkage. From equation (\ref{modelgrp}), each $(\lambda^{(m)}_{lj})^2$ is independently assigned a gamma distribution with expected value $E((\lambda^{(m)}_{lj})^2)=\frac{\alpha}{b^{(m)}_{l0}+P^{(m)T}_j\bm{b}^{(m)}_{l\cdot}}=\frac{\alpha}{b_{l0}+\sum_k\kappa^{(m)}_{jk}b^{(m)}_{lk}}$. By integrating out $(\lambda^{(m)}_{lj})^2$, the marginal prior of a non-zero loading  $a^{(m)}_{lj}$ is defined by 
\begin{equation}\label{prioA}
p(a^{(m)}_{lj})=\dfrac{\alpha+1}{2\sigma_j^{2(m)}}(b_{l0}+\sum_k\kappa^{(m)}_{jk}b^{(m)}_{lk})^{\alpha}\left(\frac{|a^{(m)}_{lj}|}{\sigma_j^{(m)}}+b_{l0}+\sum_k\kappa^{(m)}_{jk}b^{(m)}_{lk}\right)^{-\alpha-1}
\end{equation}

\begin{figure}[!ht]
\centering
\caption{Marginal prior densities of $a^{(m)}_{lj}$ for different values of $b^{(m)}_{l1}$ while other parameters are fixed: $\sigma^2=1$, $b_{l0}=0.4$, $\alpha=1$.\label{figPriorA}}
\includegraphics[height=8cm,width=10cm]{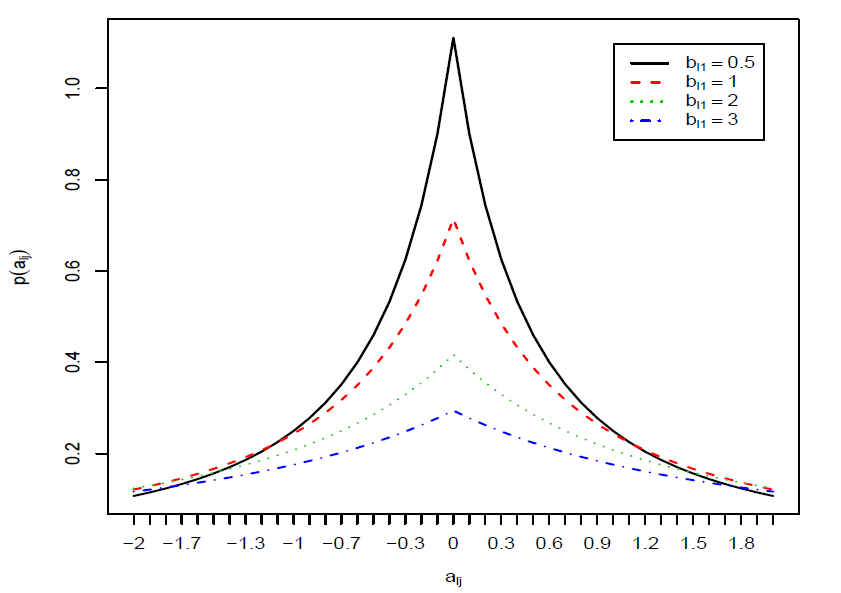}
\end{figure}

The parameter $b^{(m)}_{lk}$ is assumed to be non-negative in order to reduce the amount of shrinkage attributable to the $k$-th important group. Hence, higher values $b^{(m)}_{lk}$ give more evidence for group information by increasing the effect of features belonging to group $k$ (see Figure \ref{figPriorA} for an illustration of the  marginal prior (\ref{prioA}) of a loading).  A similar prior construction was considered by \cite{rockova2014} to incorporate grouping information in the context of regression models. 
 
\begin{figure}[H]
\caption{Contour plots of the marginal prior density of $\lambda^2_1$ and $\lambda^2_2$ for 4 different combinations of $\alpha_b$ and $\beta_b$.\label{contour}}
\begin{tabular}{cc}
\includegraphics[scale=.55]{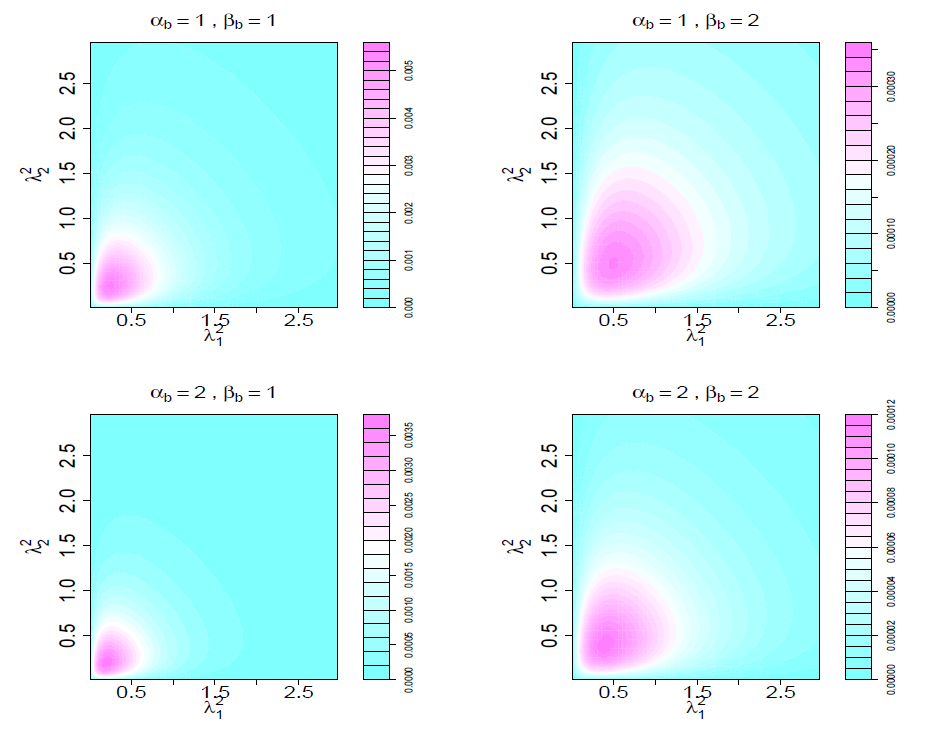}
\end{tabular}
\end{figure}

For the purpose to determine important groups that contribute to the formation of active components, we define another $(r\times K_m)$-latent binary matrix variable $\bm{R}^{(m)}=(r_{lk}^{(m)})$, where $r_{lk}^{(m)}=1$ if group (or pathway) $k$ contributes to active component $l$, and 0 otherwise. The prior of $b^{(m)}_{lk}$ is then dependent on $\bm{R}^{(m)}$ and will be defined as 
\begin{eqnarray*}
b_{lk}^{(m)}|r^{(m)}_{lk}&\sim& (1-r^{(m)}_{lk})\delta_0+r^{(m)}_{lk}\Gamma(\alpha_b,\beta_b)\nonumber\\
r^{(m)}_{lk}|q_r&\sim& \text{Bernoulli}(q_r)
\end{eqnarray*}
We finally assume that the intercept $b_{l0}^{(m)}$ follows also a Gamma $\Gamma(\alpha_0,\beta_b)$. 
Correlations between significant features within the same group are captured through the prior of  the penalty parameter $\lambda^{(m)}_{lj}$ (see prior (\ref{modelgrp})). For instance, if features 1 and 2 both belong to the same group, the joint marginal distribution of $(\lambda^2_1,\lambda^2_2)$ (after integrating out $b_0+b_1$) is
$$p(\lambda^2_1,\lambda^2_2)\propto g(\lambda^2_1,\lambda^2_2)=(\lambda^2_1\lambda^2_2)^{\alpha-1}(\beta_b+\lambda^2_1+\lambda^2_2)^{-(2\alpha+\alpha_0+\alpha_b)}. $$
Figure \ref{contour} draws the contour plots of $g(\lambda^2_1,\lambda^2_2)$ for four combinations of $(\alpha_b,\beta_b)=(1,1)$, $(1,2),(2,1),(2,2)$  with $\alpha_0=1$ and $\alpha=2$. As we would expect, as $\beta_b$ increases and/or $\alpha_b$ decreases, $\lambda^2_1$ and $\lambda^2_2$ tend to have a stronger correlation, translating to a higher probability to have similar values. 

 	\section{Posterior inference and prediction\label{postinfe}}
For posterior inference, our primary interest is in the estimation of the active component and feature selection indicator variables $\bm{\gamma}^{(m)}$, $\bm{\eta}^{(m)}$, and the loadings $\bm{A}^{(m)}=(\bm{a}^{(m)}_{lj})$. We implement a Markov Chain Monte Carlo (MCMC) algorithm for posterior inference that employs a  partially collapsed Gibbs sampling \citep{David2008} to sample   the binary matrices and the loadings. In fact, to sample $\bm{\gamma}^{(m)}$, $\bm{\eta}^{(m)}$ given the other parameters, we integrate out all the loadings $\bm{A}^{(m)}$, and we use a Metropolis-Hastings step. We then sample the loadings from its full conditional distributions. In Section \ref{postinfe2} in the web supplementary material,  we present  the  conditional distributions  used to update all the parameters. We outline the steps of our MCMC algorithm hereafter.
\begin{enumerate}
    \item Initialize the parameters $\tau_{lj}^{(2m)}$ with 1, $b^{(m)}_{l0}$ with 0.1, and the other parameters by their prior distributions. 
    \item Sample $(\bm{\gamma}^{(m)}$, $\bm{\eta}^{(m)})$ using a Metropolis-Hastings step as described in web supplementary material Section \ref{A1}.
    \item Sample the loading matrices $\bm{A}^{(m)}$, $\sigma^{2(m)}_j$, $\tau_{lj}^{(2m)}$ and $\lambda_{lj}^{(2m)}$ form their full conditionals distributions as defined in Sections \ref{A2}-\ref{A4} of the web supplementary material.
    \item Sample $r_{lk}^{(m)}$ and $b_{lk}^{(m)}$ simultaneously using a Metropolis-Hastings step as described in Section \ref{A5} of the web supplementary material.
\end{enumerate}
\textbf{Prediction:} Given a model $\{\bm{\gamma}^{(m)}$, $\bm{\eta}^{(m)},m=1,...,M\}$, 
 we estimate the loadings $\bm{A}$ for $m=0,1,...,M+1$ as $\hat{\bm{a}}_{\cdot j (\eta)}^{(m)}=\hat{\sigma}^{2(m)}_j(\bar{\bm{U}}^T_{(\gamma)}\bar{\bm{U}}_{(\gamma)}+I_{n_{\gamma}})^{-1}\bar{\bm{U}}^T_{(\gamma)}\bm{x}_{\cdot j}^{(m)}$, the posterior mode of $\bm{A}_{(\eta)}$, where $\hat{\sigma}^{2(m)}_j$ and $\bar{\bm{U}}_{(\gamma)}$  are the means of the values sampled in the course of the MCMC algorithm. Let $\bm{X}^{(m)}_{\text{new}}$, $m=1,...,M$ be the vector of features of a new individual. The latent value $\bm{U}$ of a new individual for a given model  $\{\bm{\gamma}^{(m)}$, $\bm{\eta}^{(m)},m=1,...,M\}$, is estimated as $\hat{\bm{U}}_{\text{new},(\gamma)}=(\bm{\hat{A}}_{(\eta)}D(\bm{\hat{\sigma}}^{-2})\bm{\hat{A}}_{(\eta)}^T+I_r)^{\text{-}1}\bm{\hat{A}}_{(\eta)}D(\bm{\hat{\sigma}}^{-2})\bm{x}_{\text{new}\cdot}$, where $\bm{x}_{\text{new}\cdot}=(\bm{x}^{(1)}_{\text{new}\cdot},...,\bm{x}^{(M)}_{\text{new},\cdot},\bm{x}^{(M+1)}_{\text{new},\cdot})$, $D(\bm{\hat{\sigma}}^{-2})$ is the diagonal matrix with elements $\{\hat{\sigma}^{2(m)}_j,m=1,..,M+1,j=1,...,p_m \}$ on the diagonal, and $\bm{\hat{A}}_{(\eta)}=(\bm{\hat{A}}_{(\eta)}^{(1)},...,\bm{\hat{A}}_{(\eta)}^{(M)},\bm{\hat{A}}_{(\eta)}^{(M+1)})$. Hence the predictive response is computed using a Bayesian model averaging approach \citep{BMA1999} as follows $\hat{y}_{\text{new}}=\hat{\alpha}_0+\sum_{\bm{N},\bm{\gamma}}\hat{\bm{U}}_{\text{new},(\gamma)}\bm{\hat{A}}^{(0)}_{(\eta)}p(\bm{\gamma},\bm{N}|\bar{\bm{U}}_{(\gamma)},\bm{X})$ where $\hat{\alpha}_0$ is the posterior mean of the intercept $\alpha_0$.
  \subsection*{Hyperparameter settings}
 For both simulations and observed data analysis, we set  the hyperparameters as follows. We set $r=4$ as the maximum number of active components. We note that a larger number can be set as our approach is able to select active components for each data through the indicators $\bm{\gamma^{(m)}}$. We present in Tables \ref{sensi3} and \ref{sensi4} in the web supplementary material a  sensitivity analysis for the parameter $r$. We obtain similar results with larger $r$.   We set $q_{\eta}=0.05$, the prior probability to select features for a specific component as we expect that a small number of features are active in every component (refer to Tables \ref{sensi1} and \ref{sensi2} in web supplementary material  for a sensitivity analysis of $q_{\eta}$). 
 We set $a=b=1$, hyperparameters of $q^{(m)}$ (prior probability to select a feature) and $q_r$ (prior probability to select a group) which follow each a beta distribution $(a,b)$.  We set $a_0=b_0=0.01$ to have a vague gamma prior on $\sigma_{j}^{2(m)}$. We set $\alpha=1$, the scale hyperparameter of $\lambda^{2(m)}_{lj}$.  Last, we set $\alpha_b=\beta_b=1$, hyperparameters of group effects $b_{lk}^{(m)}$ (when $r_{lk}^{(m)}=1$) that follow a gamma distribution $(\alpha_b,\beta_b)$.

 \section{Simulation studies}\label{simul}
 We consider three main scenarios to assess the performance of the proposed methods. In each scenario, there are two data types $\bm{X}^{(1)} \in \Re^{n \times p_1}$ and $\bm{X}^{(2)} \in \Re^{n \times p_2}$ simulated  according to equation (1), and a single continuous response $\bm{y} \in \Re^{n \times 1}$. We simulate each scenario to have $r=4$ latent components. The scenarios differ by how many components are shared across data types. In the first scenario, the four latent components are shared across all two data types. The first two shared components affect the response. In the second scenario, none of the $r$ components is shared across the two data types. Components 1 and 2 are unique to data type 1, and components 3 and 4 are unique to data type 2. Components 1 and 3 are associated with the response in this scenario.  In the third scenario, two of the components (1 and 2) are shared, component 3 is unique to data type 1, and component 4 is unique to data type 2. The response is associated with components 1, 3, 4. We partition the covariance matrix in each data type into signal and noise; signal contain variables that are correlated and contribute to the response, while noise variables are uncorrelated and unimportant. In each scenario, we generate 20 Monte Carlo data sets for each data type.

 \subsection{Scenario One: Each component is shared across all data types}
\begin{figure}[h]
	\centering
		\caption{Loadings $\bm{A}^{(1)}$ for settings 1, 2, 4 and 5. \label{imagesim}}
	\resizebox{\textwidth}{!}{%
		\begin{tabular}{cccc}
			\includegraphics[scale=.2]{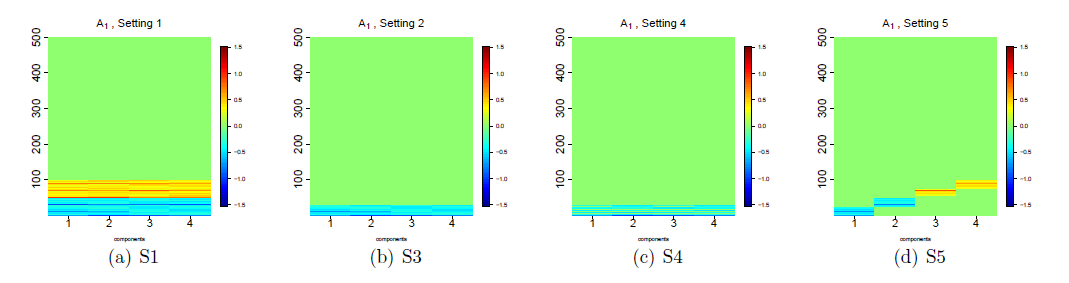}
			\\
		\end{tabular}
	}
\end{figure}

This Scenario is assumed in standard CCA methods for high dimensional data integration. 
We simulate two data types $(\bm{X}^{(1)},\bm{X}^{(2)})$, $\bm{X}^{(1)} \in \Re^{n \times p_1}$, and $\bm{X}^{(2)} \in \Re^{n \times p_2}$, $(n,p_1,p_2)=(200, 500, 500)$ according to equation (\ref{model1}) as follows. Without loss of generality, we let the first $100$ variables form the networks or groups in $\bm{X}^{(1)}$ and $\bm{X}^{(2)}$. Within each data type there are $10$ main variables, each connected to $9$ variables. 
The resulting network has $100$ variables and edges, and $p_1-100$ and $p_2-100$ singletons in $\bm{X}^{(1)}$ and $\bm{X}^{(2)}$ respectively. 
The network structure in each data type is captured by the covariance matrices
\begin{small}$\bm{\Psi}^{(1)} = \left(
\begin{array}{cc}
\bar{\bm{\Psi}}_{100 \times 100} & \bf{0} \nonumber\\
\bf{0} & \mathbf{I}_{p_1-100} \nonumber\
\end{array} \right),~~
\bm{\Psi}^{(2)} = \left(
\begin{array}{cc}
\bar{\bm{\Psi}}_{100 \times 100} & \bf{0} \nonumber\\
\bf{0} & \mathbf{I}_{p_2-100} \nonumber\
\end{array} \right).
$ \end{small}
Here, $\bar{\bm{\Psi}}$ is block diagonal with $10$ blocks of size $10$, between-block correlation $0$ and within each block there is a $9 \times 9$ compound symmetric submatrix with correlation $0.49$ describing the correlation structure of the connected variables. The correlation between a main and a connecting variable is $0.7$. We let the $p_1-100$ and $p_2-100$ singletons form one group that do not contribute to the outcome and association between $\bm{X}^{(1)}$ and $\bm{X}^{(2)}$. $\bm{A}^{(1)} \in \Re ^{r \times p_1}$ and $\bm{A}^{(2)} \in \Re ^{r \times p_2}$ are pre-specified true sparse factor loading matrices that describe the variables that are important and associated with the outcome. We consider different pre-specifications of $\bm{A}^{(1)} $ and $\bm{A}^{(2)}$ to capture the relationship between networks, and  variables and networks that are associated with the outcome. Figures \ref{imagesim}(a)-\ref{imagesim}(d) are image plots of loadings, $\bm{A}^{(1)}$, for settings 1, 2, 4 and 5. \\

\noindent \textit{\normalsize Setting 1: All groups contribute to correlation between $\bm{X}^{(1)}$ and $\bm{X}^{(2)}$}. In this scenario, we let all $10$ groups in $\bm{X}^{(1)}$ and $\bm{X}^{(2)}$ contribute to the association between $\bm{X}^{(1)}$ and $\bm{X}^{(2)}$. 
We generate the first 100 entries of $\bm{A}^{(1)} \in \Re^{r \times 100}$ and $\bm{A}^{(2)} \in \Re^{r \times 100}$ for each $r$ as independent and identically distributed (\textit{i.i.d.}) random samples from the uniform distribution on the set $[-0.5,-0.3] \cup [0.3, 0.5]$, but for the 10 main variables, we multiply the effect size by 2. The remaining $p_1-100$ entries in $\bm{A}^{(1)}$ for each $r$ is set to 0. Similarly for $\bm{A}^{(1)}$. The entries in $\bm{U}^{n \times r}$ are randomly generated from $N(0,1)$. The data matrices $\bm{X}^{(1)}$, $\bm{X}^{(2)}$, and $\bm{y}$ are generated, respectively, as 
\[ \bm{X}^{(1)}=\bm{U}\bm{A}^{(1)} + \bm{E}^{(1)}, ~~~~ \bm{X}^{(2)}=\bm{U}\bm{A}^{(2)} + \bm{E}^{(2)}, ~~~~\bm{y} = \bm{U}\bm{a} + \bm{e}\]
Here, we set $r=4$, so $\bm{U}$ represents observation from four latent components. We set $\bm{a}=[1~ 1~ 0~ 0]$ so that the response is only related to the first two shared components. The entries in  $\bm{E}^1$,  $\bm{E}^2$, and $\bm{e}$ are all generated as $i.i.d$ random samples from $N(0,1)$. \\

\noindent \textit{\normalsize Setting 2: First three groups in $\bm{A}^{(1)} $ and $\bm{A}^{(2)}$ contribute to correlation between $\bm{X}^{(1)},\bm{X}^{(2)}$.} In this scenario, we consider the situation where only some networks contribute to the association between the two data types. We let the first  $3$ groups in $\bm{A}^{(1)}$ and $\bm{A}^{(2)}$, corresponding to the first 30 variables in $\bm{X}^{(1)}$ and $\bm{X}^{(2)}$, contribute to the association between $\bm{X}^{(1)}$ and $\bm{X}^{(2)}$. The entries in 
 $\bm{A}^{(1)} \in \Re^{r \times 30}$ and $\bm{A}^{(2)} \in \Re^{r \times 30}$ are generated following Setting 1.\\
 
\noindent \textit{\normalsize Setting 3: Some variables within the groups contribute to correlation between $\bm{X}^{(1)},\bm{X}^{(2)}$}. This scenario follows that of Setting 1 but we let only some variables within each of the 10 groups to contribute to the association between $\bm{X}^{(1)}$ and $\bm{X}^{(2)}$. 
We generate the entries of $\bm{A}^{(1)} \in \Re^{r \times 100}$ and $\bm{A}^{(2)} \in \Re^{r \times 100}$ following Setting 1 but we randomly set at most 5 coefficients within each group to zero. \\

\noindent \textit{\normalsize Setting 4: Some variables and some groups in $\bm{A}^{(1)} $ and $\bm{A}^{(2)}$ contribute to correlation between $\bm{X}^{(1)},\bm{X}^{(2)}$}. This Setting follows Setting 2 but we let only some variables within 
the first three groups contribute to the association between $\bm{X}^{(1)}$ and $\bm{X}^{(2)}$. We generate the entries of $\bm{A}^{(1)} \in \Re^{r \times 30}$ and $\bm{A}^{(2)} \in \Re^{r \times 30}$ following Scenario 1 but we randomly set at most 5 coefficients within each group to  zero.\\

\noindent \textit{\normalsize Setting 5: None overlapping components and all 100 variables contribute to correlation between $\bm{X}^{(1)}$ and $\bm{X}^{(2)}$}. In Settings 1-4, we considered situations where latent components overlapped with respect to the significant variables. In this Setting, we consider none overlapping components. Specifically, data are generated according to Setting 1, but the first 25 signal variables are loaded on component 1, the next 25 variables on component 2, and so on. The last 25 signal variables are loaded on component 4. Similar to Setting 1, there are 10 groups or networks comprising of the 100 signal variables. This Setting tests our methods ability to identify all 4 components, all 10 groups, and all 100 important variables. \\

\subsubsection{Competing Methods}
We compare the performance of our methods with association only  and joint association and prediction  methods. For the association only-based methods, we consider the sparse CCA [ SELPCCA] (\cite{Safo2018}) and Fused sparse CCA [FusedCCA] (\cite{FusedCCA2018}) methods. FusedCCA incorporates prior structural information in sparse CCA by allowing variables that are connected to be selected or neglected together.  We perform FusedCCA and SELPCCA using the Matlab code the authors provide.  For FusedCCA, we assume that variables within each group are all connected since the method takes as input the connections between variables. For the joint association and prediction  methods, we consider the CCA regression (CCAReg) method  by \cite{Luo2014}. We implement CCA regression using the R package \textit{CVR} \cite{}. Once all the canonical variates  are extracted, they are used to fit a predictive model for the response $\bm{y}$, and then to compute test MSE.
\subsubsection{Evaluation Criteria}
To evaluate the effectiveness of the methods,  we fit the models using the training data and then compare different methods based on their performance in feature selection and prediction on an independently generated test data of the same size. To evaluate feature selection, we use the number of predictors in the true active set that are not selected (false negatives),  the number of noise predictors that are selected (false positives), and F-score (harmonic mean).  To evaluate prediction, we use the mean square error obtained from the testing set. We use 20 Monte Carlo datasets and report averages and standard errors. 

\subsubsection{Simulation results}
In Table \ref{res1}, we report averages of the evaluation measures for Scenario One. We first compare our method with incorporation of network information (BIPNet) and without using the network information (BIP). In Settings 1-4, both methods identify one component as important and related to the response. This is not surprising since all 100 important variables on each component overlap in this setting, resulting in our method identifying only one of these components. In Setting 5 where 
important variables did not overlap on the four latent components, the proposed methods identify at least two components 13 times out of 20 simulated data sets. We note that  in general, conmpared to BIP,  BIPNet has lower false negatives, lower positive rates, competitive or higher harmonic means, and lower MSEs. These results demonstrate the benefit of incorporating group information in feature selection and prediction. We observe a sub-optimal performance for both BIPNet and BIP in Settings 4 and 5, more so in Setting 5 where  important variables do not overlap on the latent components.  We also assess the group selection performance of BIPnet as follows. We compute AUC by using marginal posterior probabilities of group membership  $r_{lk}$'s. Table \ref{grpsel} shows that most of groups in settings 1, 2 and 5 had been identified correctly. However, the method had difficulties selecting groups in cases where only a few number of features  contribute to the association between data types (settings 3 and 4).

We next compare the proposed methods to the other existing methods. For the other methods, we estimate both the first and the first four canonical variates. We report the results from the first canonical variates here, and the results for the first four canonical covariates in the online supplementary material, Section \ref{moresim}. For feature selection, a variable is deemed selected if it has at least one nonzero coefficient in any of the four estimated components.  We observed that for SELPCCA, CCAReg, and FusedCCA in general, the first canonical variates from each data type result in lower false negatives, lower false positives, higher harmonic means, and competitive MSE for Settings 1-4 compared to results from the first four canonical variates (see web supplementary materials, Section \ref{moresim}). In Setting 5 where signal variables do not overlap on the four true latent components, we expect the first four canonical variates to have better feature selection and prediction results, and that is what we observe in Table \ref{res1}.
Across Settings 1-4, BIP and BIPNet had lower false negatives, lower or comparable false positives, higher harmonic means, and competitive MSEs when compared to two step methods: association followed by prediction. The feature selection performance of the proposed methods was also better when compared to CCAReg, a joint association and prediction method. CCAReg had  lower MSEs in general when compared to the proposed and the other methods. BIPNet and BIP had higher false negatives in Setting 5 compared to Settings 1-4, nevertheless  lower than CCAReg and SELPCCA. 

These simulation results suggest that incorporating group information in joint association and prediction methods result in better performance. Although the performance of the proposed methods are superior than existing methods in most Settings, our findings suggest that the proposed methods tend to be better at selecting most of the true signals in the settings where latent components overlap with regards to important variables. In non overlapping settings, our methods tend to select some of the true signals (resulting in higher false negatives). However, the proposed methods tend to have lower false positives, which suggest that we are better at not selecting  noise variables. This feature is especially appealing for high-dimensional data problems where there tend to be more noise variables.

 \subsection{Scenario Two: None of the component is shared across data types} We consider the scenario where there is no shared component across the two data types. This scenario is similar to standard factor analysis or principal component analysis for each data type independently, and accounts only for the individual variations explained by each data type. 
The entries in $\bm{U}$ are generated from $N(0,1)$. The data matrices are generated, respectively, as 
\[ \bm{X}^{(1)}=\bm{U}_{1,2}\bm{A}^{(1)}_{1,2} + \bm{E}^{(1)}, ~~~~ \bm{X}^{(2)}=\bm{U}_{3,4}\bm{A}^{(2)}_{3,4} + \bm{E}^{(2)}, ~~~~\bm{y} = \bm{U}\bm{a} + \bm{e}.\]
Here, $r=4$, $\bm{U}_{1,2}=(\bm{u}_1, \bm{u}_2)$ is a matrix of the first two latent components in $\bm{U}$. Similarly for $\bm{U}_{3,4}$. We consider two cases, overlapping and none overlapping components.
In the overlapping case, the true sparse loading matrix $\bm{A}^{(1)}_{1,2}$  has 100 signal variables loaded on the first two components; there are fifty variables on each component. This is also true for $\bm{A}^{(2)}_{3,4}$. In the none overlapping case, the 100 signal variables in $\bm{A}^{(1)}_{1,2}$ (similarly $\bm{A}^{(1)}_{3,4}$)  do not overlap; there are 50 variables in each component.  We set $\bm{a}=[1~ 0~1~ 0]$ so that the response is only related to the first and third components. The entries in  $\bm{E}^1$,  $\bm{E}^2$, and $\bm{e}$ are all generated as $i.i.d$ random samples from $N(0,1).$
\subsubsection{Results}
Table \ref{sim:sce2} shows the averages of the evaluation measures for Scenario Two. As before, we  estimate both the first and the first four canonical variates for the other methods, and report results using subscripts 1 and 4 respectively. From Table \ref{sim:sce2}, compared to SELPCCA, an association only method that does not utilize prior information, BIPNet and BIP have lower false negatives, lower false positives, higher harmonic means, and lower MSEs in the overlapping case. This findings hold true, generally, when compared to FusedCCA, an association only method that considers variable-variable interactions in sparse CCA. The proposed methods also yield better results in both feature selection and prediction when compared to CCAReg, a joint association and prediction method. The results in the non overlapping case in this Scenario are consistent with results from Setting 5 in Scenario One. 
For the overlapping case, our methods were able to identify the first (i.e., $\bm{A}^{(1)}_{1,2}$) and last two (i.e., $\bm{A}^{(2)}_{3,4}$) components  17 and 18 times respectively out of 20 simulated data sets (i.e 85\% and 90\%). For the none overlapping case, it was respectively 75\%, 65\%, 85\% and 65\% for components 1, 2, 3 and 4.

\subsection{Scenario Three: Some shared components and individual variations} 
We consider the scenario where some components are shared across the two data types, and there are both shared and individual components predicting the outcome. As before, the entries in $\bm{U}$ are generated from $N(0,1)$. The data matrices are generated, respectively, as 
\[ \bm{X}^{(1)}=\bm{U}_{1,2}\bm{A}^{(1)}_{1,2} + \bm{U}_{3}\bm{A}^{(1)}_{3}+ \bm{E}^{(1)}, ~~~~ \bm{X}^{(2)}=\bm{U}_{1,2}\bm{A}^{(2)}_{1,2} + \bm{U}_{4}\bm{A}^{(2)}_{4} + \bm{E}^{(2)}, ~~~~\bm{y} = \bm{U}\bm{a} + \bm{e}.\]
Here, $r=4$, $\bm{U}=(\bm{u}_1, \bm{u}_2,  \bm{u}_3,  \bm{u}_4)$, $\bm{U}_{1,2}=(\bm{u}_1, \bm{u}_2)$ is a matrix of the first two latent components in $\bm{U}$ that are shared across the two data types.  $\bm{U}_{3} = (\bm{u}_3)$ is a vector of the third latent component in $\bm{U}$ that is unique to data type $\bm{X}^{(1)}$. Similarly, $\bm{U}_{4} = (\bm{u}_4)$ is unique to data type $\bm{X}^{(2)}$. We consider two cases, overlapping shared components and none overlapping components. In the overlapping case, the true sparse loading matrices $\bm{A}^{(1)}$ have  50 signal variables (corresponding to the first 50 variables) loaded on the first two components (denoted by $\bm{A}^{(1)}_{1,2}$), 50 signal variables on the third component ($\bm{A}^{(1)}_{3}$), and no signal variable on the fourth component. This is also true for $\bm{A}^{(2)}$  but there are 50 signal variables loaded on the fourth component ($\bm{A}^{(2)}_{4}$) and no signal variables on the third component. In the none overlapping case, the 50 signal variables in $\bm{A}^{(1)}_{1,2}$ do not overlap; the first 25 variables in the first component are important, and variables 26 to 50 are important in the second component. We set $\bm{a}=[1~ 0~1~ 1]$ so that the response is only related to the first latent component that is shared, and the individual latent components that is unique to the data types. The entries in  $\bm{E}^1$,  $\bm{E}^2$, and $\bm{e}$ are all generated as $i.i.d$ random samples from $N(0,1).$

\subsubsection{Results}
This scenario tests the proposed methods ability to identify both shared and individual components, as well  as important variables that are loaded on these components. Table \ref{sim:sce3} shows the averages of the evaluation measures for Scenario Three. As before, we report the first and the first four canonical variates using subscripts 1 and 4 respectively. Similar to Setting 5 in Scenario One, the results from the first four canonical variates are generally better than that from the first canonical variates. We compare our methods to these results. From Table \ref{sim:sce3}, we observe that the proposed methods have lower false negatives, lower positives, and higher harmonic means compared to the other methods. The prediction performances for the proposed methods are sub-optimal.  Consistent with results from the other Scenarios, the proposed methods have higher false negatives in the none overlapping case, nevertheless better than CCAReg and SELPCCA. Meanwhile, the false positives are consistent in both overlapping and non overlapping cases, which again indicate that the methods are able to ignore the noise variables while selecting some of the true signals. 
In the overlapping case, our methods were able to detect components 1 and 2 for all the 20 simulated data sets. Moreover, components 3 and 4 were identified 15 and 10 times respectively. On the other hand, for the none overlapping case, components 1, 2, 3 and 4 were identified 75\%, 25\%, 25\%, and 25\% respectively. 
These findings, together with that from Scenarios One and Three, underscore the benefit of considering group information in joint association and prediction methods. 

\begin{table}[h]
\centering
\scriptsize
\caption{AUCs for group selection performance in scenario 1\label{grpsel}}
\begin{tabular}{rlllll}
  \hline
 & Setting 1 & Setting 2 & Setting 3 & Setting 4 & Setting 5 \\ 
  \hline
Data type 1 & 1 (0) & 0.86 (0.04) & 0.58 (0.03) & 0.64 (0.06) & 1 (0) \\ 
  Data type 2  & 1 (0) & 0.81 (0.05) & 0.56 (0.03) & 0.62 (0.06) & 1 (0) \\ 
   \hline
\end{tabular}
\end{table}

\begin{table}[ht]
\centering
\caption{Simulation results for Scenario One: variable selection and prediction performances. FNR1; false negative rate for $\bm{X}^1$. Similar for FNR2. FPR1; false positive rate for $\bm{X}^1$. Similar for FPR2; F11 is F-measure for $\bm{X}^1$. Similar for F12; MSE is mean square error. \label{res1}}
\scriptsize
\begin{tabular}{llllllllll}
  \hline
  Method & Setting. & FNR1 & FNR2 & FPR1 & FPR2 & F11 & F12& MSE \\ 
  \hline
BIPnet & S1 & 0 (0) & 0 (0) & 0.02 (0.02) & 0 (0) & 100 (0.00) & 100 (0) & 2.09 (0.04) \\ 
  BIP & S1 & 0 (0) & 0 (0) & 0.14 (0.06) & 0.16 (0.04) & 99.73 (0.12) & 99.68 (0.08)&  2.16 (0.04) \\ 
  SELPCCA & S1 & 0 (0) &0 (0) &0.4 (.09)&0.5 (.12)& 99.21 (0.17)  &99.02 (0.23)&2.11 (0.04)  \\
   CCAReg & S1 & 90.25(1.73) &90.40 (1.87) &1.81 (3.32)&1.71 (3.35)& 15.34 (1.87) &14.98 (2.01)&2.01 (0.05)  \\  
    FusedCCA & S1 & 0.00	(0.00)&	0.00	(0.00)&	7.88	(0.76)	&10.40	(0.78)&	86.69	(1.17)&	83.05	(1.08)&		2.14	(0.04)
\\
 \hline
  BIPnet & S2 & 0 (0) & 0 (0) & 0.21 (0.15) & 0.22 (0.21) & 98.57 (0.98) & 98.67 (1.25)  &  2.15 (0.05)  \\ 
  BIP & S2 & 0 (0) & 0 (0) & 0.61 (0.28) & 0.45 (0.19) & 96.14 (1.71) & 96.98 (1.3) & 2.2 (0.05) \\ 
  SELPCCA & S2 & 22.00 (8.75) &16.33 (7.51) &.02 (.01)&.04 (.02)& 80.17 (7.81) &85.91 (6.37)&2.13 (0.04)  \\
   CCAReg & S2 & 85.83 (1.87) &86.00 (1.79) &1.54 (.63) &1.27 (.47)& 19.52 (1.77) &19.76 (1.47) &2.03 (0.04)\\
     FusedCCA&S2&0.00	(0.00)	&0.00	(0.00)&	17.14	(0.33)&	17.13	(0.39)	&42.79	(0.47)&	42.83	(0.53)&	2.21	(0.04)\\
   \hline
  BIPnet & S3 & 0 (0) & 0 (0) & 0.89 (0.11) & 1.24 (0.12) & 96.88 (0.39) & 95.69 (0.41)&2.08 (0.04)  \\ 
  BIP & S3 & 0 (0) & 0 (0) & 0.88 (0.14) & 0.73 (0.11) & 96.93 (0.47) & 97.43 (0.4)&  2.1 (0.05) \\ 
   SELPCCA & S3 & 8.50 (5.86) &8.42 (5.81) &0.16 (.06) &.14 (.05)& 92.02 (5.07) &92.20 (5.04)&2.05 (0.04) \\
    CCAReg & S3 & 83.00 (2.50) &82.92 (2.60) &2.42 (.68)&2.51 (.74)& 23.21 (2.71) &23.02 (2.87) &1.98 (0.04) \\
    FusedCCA&S3&0.00	(0.00)	&0.00	(0.00)&	12.70	(3.01)&	13.89	(3.23)&	71.51	(2.31)&	69.80	(2.43)&	2.18	(0.04)\\
  \hline
  BIPnet & S4 & 0 (0) & 0 (0) & 2.81 (0.28) & 3.07 (0.34) & 74.73 (1.9) & 73.34 (2.26)& 2.24 (0.05) \\ 
  BIP & S4 & 0 (0) & 0 (0) & 1.82 (0.34) & 1.99 (0.36) & 83.12 (2.82) & 81.91 (2.92) &2.27 (0.04) \\ 
   SELPCCA & S4 & 51.32 (9.50) &42.89 (9.42) &0.10 (0.10)&0.00 (0.0)& 54.36 (8.08) &63.42 (8.15)&2.16 (0.04)  \\
    CCAReg & S4 & 77.63	(1.87)&	78.68	(1.85)&	3.12	(0.68)	&2.71	(0.67)	&22.80	(1.15)	&23.49	(1.43)	&2.01	(0.05)\\
    FusedCCA&S4&0.00	(0.00)&	0.00	(0.00)&	19.51	(2.22)&	19.64	(2.53)&	30.39	(1.03)&	30.55	(1.07)&		2.58	(0.06)\\
  \hline
  BIPnet & S5 & 61.05 (2.82) & 59.75 (3.12) & 0.05 (0.03) & 0 (0) & 54.91 (2.78) & 56.12 (3.03)& 2.35 (0.21) \\ 
  BIP & S5 & 63 (2.65) & 60.35 (2.48) & 0.04 (0.03) & 0.06 (0.03) & 52.89 (2.9) & 55.79 (2.63)&  2.64 (0.22) \\ 
SELPCCA & S5 & 93.40(1.55)&	93.30(1.85)&0.06(0.06)&	0.00	(0.00)&	11.60	(2.52)&	11.57	(3.00)&		2.72	(0.13)\\
   CCAReg & S5 & 88.85 (1.81) & 87.95 (1.89) &1.44 (0.52)&1.40(0.50)&  19.83 (2.30) &21.51 (2.44)&2.26 (.14) \\
   FusedCCA&S5&0.00	(0.00)&	0.00	(0.00)&	3.47	(0.81)	&2.31	(0.55)	&93.89	(1.37)&	95.77	(0.97)&		2.72	(0.11)\\
  \hline
   \hline
\end{tabular}
\end{table}

\begin{table}[htbp]	
\caption{Simulation results for Scenario Two: variable selection and prediction performances. Subscripts 1 and 4 respectively denote results from the first and the first four canonical variates. FNR1; false negative rate for $\bm{X}^1$. Similar for FNR2. FPR1; false positive rate for $\bm{X}^1$. Similar for FPR2; F11 is F-measure for $\bm{X}^1$. Similar for F12; MSE is mean square error.\label{sim:sce2}}
\begin{center}
\scriptsize
 \begin{tabular}{llllllll}
   \hline
 Method & FNR1 & FNR2 & FPR1 & FPR2 & F11 & F12 &MSE\\ 
   \hline
Overlap & & \\   
BIPnet &  0 (0) & 0.05 (0.05) & 0.05 (0.03) & 0.04 (0.02) & 99.90 (0.07) & 99.90 (0.05)& 2.12(0.04) \\ 
BIP &  0 (0) & 1.25 (0.84) & 0.21 (0.06) & 0.15 (0.05) & 99.58 (0.13) & 99.04 (0.44) & 2.16 (0.06)\\ 
 CCAReg$_1$& 71.80	(5.56)	&75.55	(5.04)	&8.81	(2.85)	&11.40	(3.32)	&29.28	(3.57)	&25.05	(3.06)& 2.19 (0.04)\\
CCAReg$_4$& 74.85	(2.63)	&74.05	(2.65)	&10.20	(1.77)	&9.78	(1.68)	&28.98	(1.77)	&30.04	(1.85) & 2.32(0.06)\\
SELPCCA$_1$& 68.75(7.23)& 93.45    (1.71)  &	0.64	(0.42)&	1.69(0.42)	&	39.71	(7.21)	&13.81	(2.76)& 2.37 (0.05)\\ 
SELPCCA$_4$& 62.70(2.86)& 56.45    (3.42)  &	4.71	(0.39)&	4.26(0.47)	&	46.65	(3.10)	&52.64	(3.83)& 2.23(0.04)\\
FusedCCA$_1$& 0.00 (0.00)& 0.00 (0.00) &	6.30	(3.21)&	10.19	(3.92)&	92.18	(3.21)	&87.30	(3.61)& 2.24 (0.07)\\
FusedCCA$_4$& 0.00 (0.00)& 0.00 (0.00) &	33.06	(4.80)&	41.81	(4.36)&	64.40	(3.93)	&57.00	(2.89)&2.18 (0.04)\\
& &\\
\hline
No Overlap & & \\
BIPnet & 29.85 (6.02) & 43.85 (5.66) & 0.01 (0.01) & 0.02 (0.02) & 79.48 (4.41) & 68.76 (4.58)& 2.00(0.09) \\ 
BIP &  34.2 (8.3) & 50.35 (8.58) & 0.11 (0.05) & 0.22 (0.07) & 72.38 (7.13) & 57.57 (7.6) &2.16 (0.06)\\    
 CCAReg$_1$& 78.65	(3.91)	&76.80	(4.29)	&6.96	(2.44)	&6.31	(2.40)	&25.41	(2.72)	&27.58	(3.01)& 2.27 (0.08)\\
 CCAReg$_4$& 69.20	(2.70)	&67.95	(2.79)	&8.25	(1.72)	&8.33	(1.66)	&36.79	(1.41)	&37.85	(1.42)&1.77(0.04)\\
 SELPCCA$_1$& 87.35(1.48)& 87.90    (1.89)  &	0.36	(0.16)&	0.31(0.09)	&	21.59	(2.31)	&20.46	(2.80)& 2.44 (0.07)\\
SELPCCA$_4$& 59.35(2.81)& 57.55    (1.69)  &	4.51	(1.84)&	3.10(1.02)	&	51.08	(2.49)	&55.28	(2.32)&2.02(0.07)\\
 FusedCCA$_1$& 0.00 (0.00)& 0.00 (0.00) &	1.45	(0.65)&	5.53	(1.63)&	97.44	(1.08)	&91.35	(2.35) & 2.23 (0.07)\\
FusedCCA$_4$& 0.00 (0.00)& 0.00 (0.00) &	25.26	(3.44)&	30.21	(4.70)&	69.11	(3.16)	&66.53	(3.88)&1.84 (0.04)\\ 
\hline
\hline
\end{tabular}
	\end{center}
\end{table}

\begin{table}[htbp]	
\caption{Simulation results for Scenario Three: variable selection and prediction performances. Subscripts 1 and 4 respectively denote results from the first and the first four canonical variates. FNR1; false negative rate for $\bm{X}^1$. Similar for FNR2. FPR1; false positive rate for $\bm{X}^1$. Similar for FPR2; F11 is F-measure for $\bm{X}^1$. Similar for F12; MSE is mean square error. \label{sim:sce3}}
\begin{center}
\scriptsize
 \begin{tabular}{llllllll}
   \hline
 Method & FNR1 & FNR2 & FPR1 & FPR2 & F11 & F12 &MSE\\ 
   \hline
\textbf{Overlap} & & \\   
  BIPnet &  30.2 (4.7) & 23.25 (5.22) & 0.01 (0.01) & 0.05 (0.02) & 80.52 (3.18) & 84.78 (3.49)& 2.90 (0.14) \\ 
BIP &  31.55 (4.73) & 14.5 (5.02) & 0.12 (0.04) & 0.11 (0.05) & 79.34 (3.14) & 90.2 (3.41) & 3.26 (0.5)\\  
CCAReg$_1$&  82.30(	2.57)&82.10	(2.95)&	4.50	(1.29)&	4.25	(1.35)&	24.23	(2.52)&	24.45	(2.69)& 3.11 (0.05)\\
CCAReg$_4$&  72.60(	2.66)&74.60	(2.10)&	9.38	(1.98)&	9.40	(2.13)&	32.13	(1.49)&	30.74	(1.24)&2.38 (0.05)\\
SELPCCA$_1$& 83.55(4.42)& 93.05    (2.28)  &	0.01	(0.01)&	0(0)	&24.18	(5.61)	&11.82	(2.92)& 3.26 (0.05)\\
SELPCCA$_4$& 61.65(3.67)& 63.00    (2.07)  &	5.05	(1.94)&	2.43(0.34)	&	47.73	(1.96)	&50.07	(2.56) & 2.52 (0.07)\\
FusedCCA$_1$& 0.00 (0.00)& 0.00 (0.00) &	5.15	(0.55)&	7.00	(0.69)&	90.83	(0.89)	&87.96	(1.04)& 3.19 (0.06)\\
FusedCCA$_4$& 0.00 (0.00)& 0.00 (0.00) &	32.28	(3.54)&	24.59	(3.62)&	64.00	(3.01)	&69.97	(3.22)& 2.37 (0.04)\\
& & \\
\hline
\textbf{No Overlap} & &\\
BIPnet &  51.75 (4.76) & 43.85 (6.06) & 0.04 (0.02) & 0.05 (0.03) & 62.28 (4.52) & 67.9 (5.33) & 3.42 (0.48) \\ 
BIP &  56.7 (5.76) & 41.05 (5.67) & 0.1 (0.03) & 0.14 (0.05) & 56.14 (5.4) & 70.46 (5.07) & 3.26 (0.5) \\ 
CCAReg$_1$& 88.60(	7.92)&87.10	(2.24)&	2.67	(0.62)&	2.00	(0.52)&	17.38	(2.68)&	19.72	(2.99)&   3.00 (0.09)\\
CCAReg$_4$& 76.80(	2.61)&73.60	(2.92)&	7.90	(1.79)&	7.49	(1.93)&	28.88	(1.23)&	32.62	(1.37)& 2.06 (0.06)\\
     SELPCCA$_1$& 94.85(1.64)& 92.15    (2.09)  &	0.40	(0.40)&	0.01(0.01)	&	8.57	(2.22)	&13.30	(3.36)& 3.46 (0.08)\\
     SELPCCA$_4$& 67.90(2.39)& 64.55    (1.73)  &	2.60	(1.25)&	0.98(0.23)	&	44.53	(1.81)	&50.49	(1.98)& 2.35 (0.05)\\
      FusedCCA$_1$& 0.00 (0.00)& 0.00 (0.00) &	6.10	(1.66)&	5.11	(1.11)&	90.35	(2.21)	&91.37	(1.71)& 3.18 (0.08)\\
      FusedCCA$_4$& 0.00 (0.00)& 0.00 (0.00) &	32.13	(4.47)&	24.84	(5.27)&	64.33	(3.44)	&72.40	(4.36)& 2.50 (0.07)\\
 \\ 
\hline
\hline
\end{tabular}
	\end{center}
\end{table}

\clearpage
\section{Application to atherosclerosis disease\label{appli}}
We apply the proposed methods to analyze gene expression, genetics, and clinical data from the PHI study. The main goals of our analyses are to: 1) identify genetic variants, genes, and gene pathways that are potentially predictive of 10-year atherosclerosis disease (ASCVD), and 2) illustrate the use of the shared components or scores in discriminating subject at high- vs low-risk for developing ASCVD in 10 years. 

There were 340 patients with gene expression and SNP data, as well as clinical covariates to calculate their ASCVD score. The following clinical covariates were added as a third data set: age, gender, BMI, systolic blood pressure, low-density lipoprotein (LDL), and triglycerides. The gene expression and SNP data were each standardized to have mean zero and standard deviation one for each feature. Details for data preprocessing are provided in web supplementary material Section \ref{dataprocess}. 

To obtain the group structure of genes, we performed a network analysis using Ingenuity Pathway Analysis (IPA), a software program which can analyze the gene expression patterns using a build-in scientific literature based database (according to IPA Ingenuity Web Site, www.ingenuity.com). By mapping our set of genes with the IPA gene set, we identified $p_1=561$ genes within $K_1= 25$ gene networks. To obtain the group structure of SNPs ($p_2=413$ SNPs), we identified genes nearby SNPs on the genome using the R package \textit{bioMart}.  We found in total $K_2=31$ genes that will represent groups for the SNP data. The list of those groups for both data types are shown in the web supplementary material, Section \ref{secgroups}. 

We divided each of the data type into training  ($n=272$) and testing  ($n=68$), ran the analyses on the training data to estimate the latent scores (shared components) and loading matrices for each data, and predicted the test outcome using the testing data. For the other methods, the training data were used to identify the optimal tuning parameters and to estimate the loading matrices. We then computed the training and test scores using the combined scores from both the gene expression and SNP data. We fit a linear regression model using the scores from the training data,  predicted the test outcome using the test scores, and  computed test MSE.  

To reduce the variability in our findings due to the random split of the data, we also obtained twenty random splits, repeated the above process, and computed average test MSE and variables selected. We further attempted to assess the stability of the variables selected by considering variables that were selected at least 12 times (60\% ) out of the twenty random splits. \\

\begin{figure}[H]
	\centering
		\caption{MPPs of the 4  component indicators, $\gamma^{(m)}_{l}$, $l=1,2,3,4$, and $m=$Response (i.e. ASCVD scores), Genes, SNPs and covariates, with respect to our four proposed methods.\label{margcompo}}
	\resizebox{\textwidth}{!}{%
		\begin{tabular}{cc}
			\includegraphics[scale=0.8]{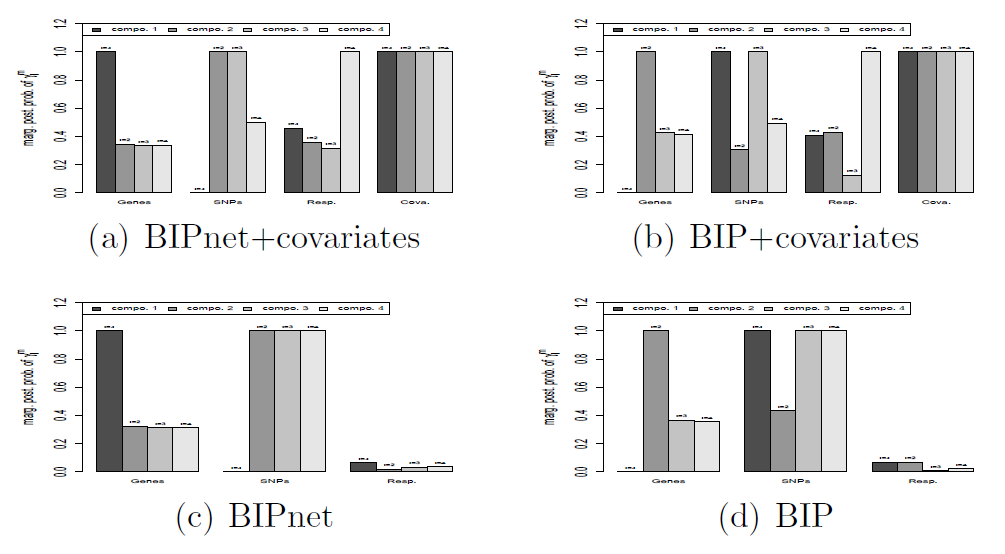}
			\\
		\end{tabular}
	}

\end{figure}

\noindent\normalsize{\textit{\textbf{Prediction error and genes and genetic variants selected}}}: 
For the purpose of illustrating the benefit of including covariates, we also considered these additional models: i) our BIPnet model, using both molecular data and clinical data (age, gender, Body mass index, Systolic blood pressure, low-density lipoprotein (LDL),  triglycerides) [BIPnet+Cov], (ii) our BIP model, using both clinical  and molecular data (BIP+ Cov), and iii) sparse PCA on stacked omics and clinical data (SPCA + Cov). For the sparse PCA, we used the method of \cite{witten2009penalized} with $l_1$-norm regularization.  We were not able to incorporate clinical data in the integrative analysis methods under comparison since they are applicable to two datasets. 
Figure \ref{margcompo} shows marginal posterior probabilities (MPP) of the 4 components for each datatype (SNPs, mRNA and response) for one random split of the data. It shows that for each of the 4 methods, only one component is active (i.e MPP$>0.5$) for gene expression, and two components for SNP data. None of the component is active for the response variable when applied to methods without clinical data (BIPnet and BIP). However, when applied to methods that incorporate covariates (BiPnet+Cov, BIP+Cov), we can obviously identify at least one active component for the response variable. This result shows the importance of incorporating clinical data in our approach.

Table \ref{real:tab1} gives the average test mean square error and numbers of genes and genetic variants identified by the methods for twenty random split of the data. For FusedCCA, we assumed that all genes within a group are connected (similarly for SNPs). We computed the first and the first four canonical variates for FusedCCA, SELPCCA, and CCAReg using the training dataset, and predicted the test ASCVD score using the combined scores from both the gene expression and SNP data.  BPINet+ Cov and BIP + Cov have better prediction performances (lower MSEs) when compared to BIPNet, BIP, and the other methods. We observe from  that BIPNet and BIPNet + Cov tend to be more stable than BIP and BIP + Cov (125 genes and 55 SNPs vs 0 genes and 54 SNPs selected at least 12 times). Together, these findings demonstrate the merit in combining molecular data, prior biological knowledge, and clinical covariates in integrative analysis and prediction methods. 

We further assessed the biological implications of the groups of genes and SNPs identified by the proposed methods. We selected (MPP$>$0.5) in total  125 genes and 55 SNPs at least 12 times out of the 20 random splits for both BIPNet and BIPNet + Cov (see Figures \ref{figure7} and \ref{figure8} in supplementary material for a random split). Networks 1 and 10 (refer to web supplementary material Table \ref{tableC2} for group description)  were identified 18 times (90\%) out of the 20 random splits of the data. Cell-To-Cell signaling and interaction that characterizes network 10  has led to novel therapeutic strategies  in cardiovascular diseases \citep{Shaw2012}. Genes PSMB10 and  UQCRQ, and SNPs \textit{rs1050152} and \textit{rs2631367} were the two top selected genes (highest averaged MPPs) and SNPs respectively across the 20 random splits.  The gene PSMB10, a member of the ubiquitin-proteasome system, is a coding gene that encodes the Proteasome subunit beta type-10 protein in humans. PSMB10 is reported to play an  essential role in maintaining cardiac protein homeostasis \citep{li2018novel} with a compromised proteasome likely contributing to the pathogenesis of cardiovascular diseases \citep{wang2015protein}. The SNPs \textit{rs1050152} and \textit{rs2631367} found  in genes SLC22A4 and SLC22A5, belong to the gene family SLC22A which has been implicated in cardiovascular diseases.  Other genes and SNPs with MPP$> .85$, on average, are reported in the web supplementary material in Tables \ref{tableC1} and \ref{tableC3}.  These could be explored for their potential roles in cardiovascular (including atherosclerosis) disease risk. \\

\begin{figure}[H]
	\centering
		\caption{Estimated latent U scores on both training and test sets for low and high CVD risks  with respect to our methods.\label{AUCresponseBIP}}
	\resizebox{\textwidth}{!}{%
		\begin{tabular}{cc}
			\includegraphics[scale=0.3]{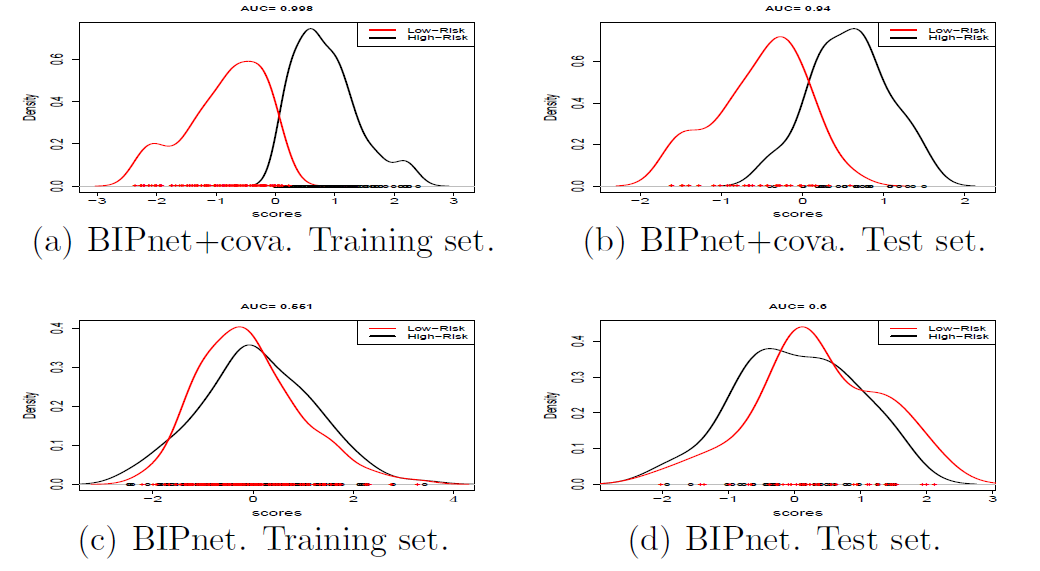} \\ 
		\end{tabular}
	}

\end{figure}

\noindent\normalsize{\textit{\textbf{Discriminating between high- vs low-risk ASCVD using the shared components:}}}
We illustrate the use of the shared components or scores in discriminating subjects at high- vs low-risk for developing ASCVD in years.  For this purpose, we dichotomized the response into two categories: low-risk and high-risk patients. ASCVD score less or equal to the median ASCVD was considered low-risk, and ASCVD score above the median ASCVD was considered high-risk. We computed the AUC  for response classification using a simple logistic regression model with the  most significant (i.e. the highest MPP for our method)  component associated with the response as covariates. For the other methods, the  AUC was calculated using the first canonical correlation variates from the combined scores.
Figure \ref{AUCresponseBIP} shows that the components selected by our methods applied to the omics and clinical data are able to discriminate clearly the two risk groups  compared to our methods applied to the omics data only. The estimated AUCs for the omics plus clinical covariates are much higher than those obtained from omics only applications (e.g AUC is 1.00 for the training set and AUC is 0.94 for the test set using BIPnet+Cov). When compared to the two-step methods CCAReg and SELPCCA, the estimated AUCs from our methods are considerably higher than the AUC's from these methods. These findings demonstrate the importance of one-step methods like we propose here, and also the merit in incorporating both clinical and molecular data in integrative analysis and prediction models.    

\begin{table}[htbp]
\begin{center}
\scriptsize
\caption{Average test mean squared error and number of variables selected by the methods for twenty random split of the data. Genes/SNPs selected $\ge 12$ are genes or snps selected at least 12 times out of twenty random splits. Subscripts 1 and 4 respectively denote results from the first and the first four canonical variates.  MSE is mean square error.\label{real:tab1}}
	\begin{small}
		\begin{tabular}{llrrrrr}
			\hline
		    Method &	 MSE 	& Average &	Average& Genes selected &SNPs selected \\
		     &	 	& \# Genes &	\# SNPs&$\ge 12$ times&$\ge 12$ times\\
			\hline
			BIPnet &1.362 (0.033)     &128.35	&113.65&125     &55	 \\
		   BIPnet + Cov &0.758 (0.125)     &122	&95.8 &125     &55  \\
		    BIP&1.363 (0.032) &68.7 &88.8&0&54 \\
		     BIP + Cov &0.766 (0.153)     &54.7	&82.5 &0      &54	 \\
		     SPCA$_1$ + Cov  & 1.367 (0.035)&0 & 5.55 & 0 & 0\\ 
		     SPCA$_4$ + Cov  & 1.382 (0.036) &23.95 & 132.60&  0& 113\\  
          	SELPCCA$_1$    & 1.373 (0.035)  	&52.35	&30.35   & 1 	&23\\ 
            SELPCCA$_4$    & 1.377 (0.036)  	&310.650&170.60   & 174  	&102\\
			CCAReg$_1$& 1.371(0.008)              &48.85	&43.55 & 0             &0 \\ 
			CCAReg$_4$& 1.382 (0.034)              &11.30	&12.85&0              &0	 \\ 
			FusedCCA$_1$&1.370 (0.032) &560.2 & 402.40 &561 &398  \\	
			FusedCCA$_4$&1.379 (0.034) &561 & 418.05 &561 &421  \\	
			\hline
			\hline
		\end{tabular}
	\end{small}
\end{center}
\end{table}

	\begin{figure}[H]
	\centering
		\caption{Estimated latent U scores on both training and test sets for low and high CVD risks  with respect to competing methods. \label{fig:score}}
	\resizebox{\textwidth}{!}{%
		\begin{tabular}{cc}		
			\includegraphics[scale=0.3]{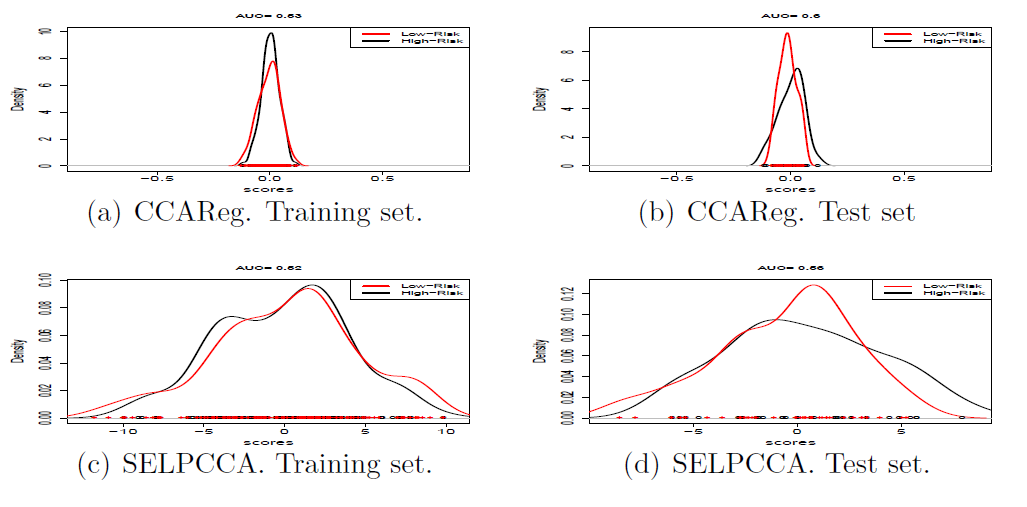}
		\end{tabular}
	}
\end{figure}
\section{Conclusion\label{conclu}}
We have presented a Bayesian hierarchical modeling framework that integrates in a unified procedure multi-omics data, a response variable, and clinical covariates. We also extended the methods to integrate grouping information of features through prior distributions. The numerical experiments  described in this manuscript show that our approach outperforms competing methods mostly in terms of variable selection performance. The data analysis demonstrated the merit of integrating omics, clinical covariates, and prior biological knowledge, while simultaneously predicting a clinical outcome. 
The identification of important genes or SNPs and their corresponding important groups ease the biological interpretation of our findings. 

Several extensions of our model are worth investigating. First, using latent variable formulations,  we can naturally extend our approach to other types of response variables such as survival time, binary or multinomial responses or mixed response outcomes (e.g. both continuous and binary). Second, the proposed method  is only applicable
to complete data and do not allow for missing values. A future project could extend the
current methods to the scenario where data are missing using Bayesian imputation methods or the approach proposed by \cite{Cheks2017} for missing blocks of data. 
\if1\blind
{
\section*{Acknowledgements}
We are grateful to the Emory Predictive Health Institute for providing
us with the gene expression, SNP, and clinical data.  Sandra Safo   is partly supported by NIH grant 1KL2TR002492\-02, and  Thierry Chekouo is partially supported by NSERC Discovery Grants number RGPIN-2019-04810.  The content is solely the responsibility of the authors and does not necessarily represent the official views of the NIH and NSERC .
}  \fi
\section*{Supplemental Material}
In the online Supplemental Materials, we provide details to sample our parameters, and present more results from simulated data, in particular, sensitivity analysis results. We provide a detailed description of the groups from each of the data types (SNPs and gene expression).  We have created an R package called \textit{BIPnet} (GNU zipped tar file)  for implementing the methods. Its source R and C codes, along with the user pdf manual are submitted  with this manuscript. The package also contains functions used to generate simulated data.  Data used for the analyses may be requested from the Emory Predictive Health Institute.

%
 				
\section*{Supplementary materials for  ``Bayesian Integrative Analysis and Prediction with Application to Atherosclerosis Cardiovascular Disease"}

\section{Posterior inference\label{postinfe2}}
We present below the  conditional distributions  used to update the parameters. 
\subsection{Sampling $\gamma^{(m)}_{l}$ and $\eta^{(m)}_{lj}$}\label{A1}	
Equation (1) in the main paper can also be written as  
$\bm{x}_{\cdot j}^{(m)}=\bm{U}_{(\gamma)}\bm{a}_{\cdot j (\eta)}^{(m)}+\bm{e}_{\cdot j}^{(m)},$ $j=1,...,p_m,$
where $\bm{a}_{\cdot j (\eta)}^{(m)}$ is the $j$-column vector of $\bm{A}^{(m)}$ with $\eta^{(m)}_{lj}=1$, and $\bm{U}_{(\gamma)}$ is a sub-matrix of $\bm{U}$ with columns $l$'s verifying $\gamma^{(m)}_{l}=1$.
After integrating $\bm{a}_{\cdot j (\eta)}^{(m)}$, we have 
\begin{eqnarray}
p(\bm{x}^{(m)}_{\cdot j}|\bm{\gamma}^{(m)},\bm{U}_{(\gamma)},\bm{\eta_{\cdot j}}^{(m)},\sigma^{2(m)}_j,\bm{\tau}_j)&=&MVN(\bm{x}_{\cdot j},\bm{0},\sigma^{2(m)}_j\bm{\Sigma}_j) \nonumber
\end{eqnarray}
where $\bm{\Sigma}_j=\bm{U}_{(\gamma)}D(\bm{\tau}_j)\bm{U}^T_{(\gamma)}+\bm{I}_n$. In what follows, we removed the superscript ``$^{(m)}$"  to simplify formulas. We have then,
\begin{eqnarray}
p(\bm{\gamma},\bm{N}|\bm{X},\bm{U},\bm{\sigma},\bm{\tau})&\propto &\prod_{j=1}^{p}p(\bm{x}_{\cdot j}|\bm{\gamma},\bm{U}_{(\gamma)},\bm{\eta_{\cdot j}},\sigma^2_j,\bm{\tau}_j)p(\bm{\eta_{\cdot j}}|\bm{\gamma})p(\bm{\gamma})\nonumber\\
&\propto &\prod_{j=1}^{p}MVN(\bm{x}_{\cdot j},\bm{0},\sigma^2_j\bm{\Sigma}_j)\prod_{l=1}^{r}p(\eta_{l j}|\gamma_l)p(\gamma_l)\nonumber\\
&=&\prod_{j=1}^{p}G_j(\bm{\gamma},\bm{\eta_{\cdot j}})\prod_{l=1}^{r}p(\gamma_l),\label{MHrho}
\end{eqnarray}
where $\bm{N}=(\eta_{lj})$ and $G_j(\bm{\gamma},\bm{\eta_{\cdot j}})=MVN(\bm{x}_{\cdot j},\bm{0},\sigma^2_j\bm{\Sigma}_j)\prod_{l=1}^{r}p(\eta_{l j}|\gamma_l)$.
To sample $(\bm{\gamma},\bm{N})$ from equation (\ref{MHrho}), we adopt a Metropolis Hasting step. The proposal distribution will be defined as 
\begin{equation}
Q(\gamma'_l,\bm{\eta'}_{l\cdot }|\gamma_l,\bm{\eta}_{\cdot l})=\gamma_l\delta_0(\gamma'_l) \delta_{\bm{0}}(\bm{\eta'}_{l\cdot })+ (1-\gamma_l)\delta_1(\gamma'_l)p(\bm{\eta'}_{l\cdot}|\bm{X},\bm{U},\gamma'_l=1) \label{propRho}              
\end{equation}
where 
$$p(\bm{\eta'}_{l \cdot }|\bm{X},\bm{U},\gamma'_l=1)=\prod_{j=1}^{p}P_{lj}^{\eta'_{lj}}(1-P_{lj})^{1-\eta'_{lj}}, \quad P_{lj}=\frac{G_j(\bm{\gamma}^1,\bm{\eta_{\cdot j}}^1)}{G_j(\bm{\gamma}^1,\bm{\eta_{\cdot j}}^1)+G_j(\bm{\gamma}^1,\bm{\eta_{\cdot j}}^0)},$$ $\bm{\gamma}^1=(\gamma_1,...,\gamma_l=1,...\gamma_r)$, $\bm{\eta_{\cdot j}}^1=(\eta_{1j},...,\eta_{lj}=1,...,\eta_{rj})$ and $\bm{\eta_{\cdot j}}^0=(\eta_{1j},...,\eta_{lj}=0,...,\eta_{rj})$. 

From  proposal (\ref{propRho}), if the current component is  not active, that is, $\gamma_l=0$, we propose to activate this group by setting $\gamma'_l=1$, and sample from the conditional distribution, $p(\bm{\eta'}_{l \cdot }|\bm{X},\bm{U},\gamma'_l=1)$, of the indicators for variable selection $\bm{\eta}_{l\cdot }$. We then determine whether to accept this
proposal or not by the Metropolis-Hastings acceptance-rejection rule. Conversely, if the
current component is active, that is, $\gamma_l=1$, we propose to make it inactive by setting
$\gamma'_l=0$ and setting all the indicators $\bm{\eta'}_{l \cdot }$ within this component to zero. Again, we determine whether to accept this proposal or not by the acceptance-rejection rule. When $m=0$, we will always assume $\eta_{l1}=1$ and only $\gamma^{0}_l$ is updated.

\subsection{Sampling $\sigma_{j}^{2(m)}$, $\tau_{lj}^{2(m)}$ and  $\lambda^{2(m)}_{lj}$}\label{A2}	
We remove the superscript $^{(m)}$ to simplify the notations. 
We easily show that the full conditional distributions of $\sigma_{j}^{2}$'s are obtained as
$$p(\sigma_{j}^{2}|\cdot)= IG(\sigma_{j}^{2};a_0+\frac{n}{2},\frac{1}{2}\bm{x}^T_{\cdot j}\Sigma^{-1}_j\bm{x}_{\cdot j}+b_0).$$
Also, when  $\eta_{lj}=1$, the full conditional distributions of $\tau_{lj}^{2}$ and  $\lambda^{2}_{lj}$ are respectively the inverse of inverse Gaussian distribution and a gamma distribution. More specifically,
\begin{eqnarray}
\frac{1}{\tau_{lj}^{2}}|\cdot&\sim & \text{InverseGaussian}(\mu',\lambda'^2)\nonumber\\
\text{and } \lambda^2_{lj}|\cdot&\sim &\Gamma (\alpha+1, b_{l0}+P_j^T\bm{b}_l+\tau^2_{lj})\nonumber
\end{eqnarray}
$$\text{where }
\mu'=\sqrt{\frac{2\lambda^2_{lj}\sigma^2_j}{A_{lj}^2}} \text{ and } \lambda'^2=2\lambda^2_{lj}$$

Otherwise, we update them from these  distributions (pseudo-priors)
\begin{equation}
\tau^2_{lj}|\lambda^2_{lj}\sim\text{Exp}(\lambda^2_{lj}) \text{ and }
\lambda^2_{lj}|\bm{b}_{l\cdot}\sim\Gamma(\alpha,b_{l0}). \nonumber
\end{equation}
\subsection{Sampling $\bm{A}$}\label{A3}
Full conditional distributions of  $\bm{a}_{\cdot j (\eta)}^{(m)}$ are obtained as 
\begin{equation}
p(\bm{a}_{\cdot j (\eta)}^{(m)}|..)=Normal(\bm{a}_{\cdot j (\eta)}^{(m)};\mu_{a},\Sigma_{a})   \nonumber
\end{equation}
where $\mu_{a}=\Sigma_{a}\bm{U}^T_{(\gamma)}\bm{x}_{\cdot j}^{(m)}$, $\Sigma^{-1}_{a}=\sigma^{-2(m)}_j(\bm{U}^T_{(\gamma)}\bm{U}_{(\gamma)}+I_{n_{\gamma}})$ and $n_{\gamma}$ is the number of active components for $m$. We also have $a^{(m)}_{lj}=0$ if $\eta^{(m)}_{lj}=0$. 
\subsection{Sampling  $\bm{U}$} \label{A4}
Let denote $p=p_0+p_{M+1}\sum_{m=1}^{M}\sum_{j=1}^{p_m}\eta^{(m)}_{lj}$, the total number of features included in the model (including the outcome and the set of covariates). We can write also $\bm{x}_{i\cdot}^{(m)}=\bm{A}^{(m)T}\bm{u}_{i\cdot}+\bm{e}_{i\cdot}^{(m)}$, $i=1,...,n$ and let $\bm{x}_{i\cdot}=(y_i,\bm{x}_{i\cdot}^{(1)},\ldots,\bm{x}_{i\cdot}^{(M)})$ be the $p$-vector of expression of sample $i$ and $A_{(\eta)}=(\bm{A}^{(0)},\bm{A}_{(\eta)}^{(1)},\ldots, \bm{A}_{(\eta)}^{(M)})$ be the $r\times p$ matrix of loads. 
\begin{eqnarray}
P(\bm{u}_{i\cdot}|\bm{x}_{i\cdot},\bm{A}_{(\eta)})&\propto&	P(\bm{x}_{i\cdot}|\bm{u}_{i\cdot},\bm{A}_{(\eta)})P(\bm{u}_{i\cdot})\nonumber\\
&=&\exp\{\text{-}\frac{1}{2}[\bm{x}_{i\cdot}-\bm{A}_{(\eta)}^{T}\bm{u}_{i\cdot}]^TD(\bm{\sigma}^{-2})[\bm{x}_{i\cdot}-\bm{A}_{(\eta)}^{T}\bm{u}_{i\cdot}]\}\exp\{\text{-}\frac{1}{2}\bm{u}_{i\cdot}^T\bm{u}_{i\cdot}\}\nonumber\\
&=& \exp\{\text{-}\frac{1}{2}[\bm{u}_{i\cdot}-\mu_i]^T\Sigma^{-1}_{u_i}[\bm{u}_{i\cdot}-\mu_i]\}\nonumber
\end{eqnarray}
where $\bm{\sigma}$ is the vector $(\sigma^{(0)},\bm{\sigma}^{(1)},\ldots,\bm{\sigma}^{(M)})$. Hence, $\bm{u}_{i\cdot}$ is a multivariate normal distribution with mean $\mu_{u_i}=\Sigma_{u_i}\bm{A}_{(\eta)}D(\bm{\sigma}^{-2})\bm{x}_{i\cdot}$  and covariance matrix $\Sigma_{u_i}=(A_{(\eta)}D(\bm{\sigma}^{-2})A_{(\eta)}^T+I_r)^{\text{-}1}$.

\subsection{Sampling  $(b^{(m)}_{lk},r^{(m)}_{lk})$, $l=1,...,r$}\label{A5}

\begin{eqnarray}
P(b_{lk},r_{lk}|\cdot)&\propto&\prod_{\substack{j=1\\ \eta_{lj}=1}}^{p}P(\lambda^2_{lj}|\bm{b}_l)P(b_{lk}|r_{lk})P(r_{lk})\nonumber\\
&\propto&\prod_{\substack{j=1\\ \eta_{lj}=1}}^{p_m}(b_{l0}+\sum_{k=1}^{K}\kappa_{jk}r_{lk}b_{lk})^{\alpha}\exp\{-\kappa_{jk}\lambda_{lj}r_{lk}b_{lk}\}\times\nonumber\\
&\times& [(1-r_{lk})\delta_0+r_{lk}\frac{\beta_0^{\alpha_0}}{\Gamma(\alpha_0)}b_{lk}^{\alpha_0-1}\exp(-\beta_0b_{lk})]q_r^{r_{lk}}(1-q_r)^{1-r_{lk}}\nonumber
\end{eqnarray}
where $P_j=(\kappa_{jk})_K$ and $\kappa_{jk}=1$ if $j$ belongs to group $k$. 
We will sample $(b_{lk},r_{lk})$ using this proposal (similar to \cite{Gottardo2008})
\begin{equation} 						Q(b'_{lk},r'_{lk}|b_{lk},r_{lk})=r_{lk}\delta_0(b'_{lk})\delta_0(r'_{lk})+(1-r_{lk})\delta_1(r'_{lk}) \Gamma(b'_{lk};\alpha^{\star},\beta^{\star})\label{propR} \nonumber
\end{equation}
where $\Gamma(b'_{lk};\alpha^{\star},\beta^{\star})$ is the density of a gamma distribution with parameters $\alpha^{\star}$ and $\beta^{\star}$ to specify. From proposal (\ref{propR}), if the current pathway is selected in component $l$, i.e, $r_{lk}=1$, we propose $r'_{lk}=0$ and $b'_{lk}=0$. Otherwise, if $r_{lk}=0$, we propose $r'_{lk}=1$ and $b'_{lk}\sim \Gamma(\alpha^{\star},\beta^{\star})$.

\section{More simulation results}\label{moresim}

Table \ref{sim:sce1b} 	shows complete simulation results of the competitive methods applied with both $2$ and $4$ shared components.	

\begin{scriptsize}
	\begin{longtable}{llllllllll}
		\caption{Simulation results for Scenario One: variable selection and prediction performances. Subscripts 1 and 4 respectively denote results from the first and the first four canonical variates. FNR1; false negative rate for $\bm{X}^1$. Similar for FNR2. FPR1; false positive rate for $\bm{X}^1$. Similar for FPR2; F11 is F-measure for $\bm{X}^1$. Similar for F12; MSE is mean square error. FusedCCA$_4$: method Fused CCA applied with 4 canonical variates.  FusedCCA$_1$: method Fused CCA applied with  one canonical variate \label{sim:sce1b}}\\
		\hline
		\endfirsthead
		\endhead
		\hline\multicolumn{2}{l|}{Continued on next page} \\ 
		\endfoot
		\hline
		\endlastfoot
		
		Method & Setting & FNR1 & FNR2 & FPR1 & FPR2 & F11 & F12& MSE \\ 
		\hline
		BIPnet & S1 & 0 (0) & 0 (0) & 0.02 (0.02) & 0 (0) & 99.95 (0.05) & 100 (0) & 2.09 (0.04) \\ 
		BIP & S1 & 0 (0) & 0 (0) & 0.2 (0.05) & 0.12 (0.04) & 99.6 (0.1) & 99.75 (0.08)&  2.16 (0.04) \\ 
		SELPCCA$_1$ & S1 & 0 (0) &0 (0) &0.4 (.09)&0.5 (.12)& 99.21 (0.17)  &99.02 (0.23)&2.11 (0.04)  \\
		SELPCCA$_4$ &S1& 31.35(	5.24)&	26.60	(5.30)	&13.00	(3.85)&	12.38	(3.51)&	63.28	(3.27)&	67.07	(4.23)&	2.09	(0.05)\\ 		
		CCAReg$_1$ & S1 & 90.25(1.73) &90.40 (1.87) &1.81 (3.32)&1.71 (3.35)& 15.34 (1.87) &14.98 (2.01)&2.01 (0.05)  \\ 
		CCAReg$_4$& S1&77.80	(4.32)&	78.50	(4.51)&	13.68	(3.32)&	14.05	(03.35)&	20.27	(3.05)&	19.34	(2.74)&	2.22	(0.05)\\ 				
		FusedCCA$_1$ & S1 & 0.00	(0.00)&	0.00	(0.00)&	7.88	(0.76)	&10.40	(0.78)&	86.69	(1.17)&	83.05	(1.08)&		2.14	(0.04)	\\
		FusedCCA$_4$&S1&0.00	(0.00)&	0.00	(0.00)&	33.63	(4.11)&	27.59	(3.16)&	62.56	(3.06)&	66.38	(2.55)&	2.19	(0.04)\\ 		
		\hline
		BIPnet & S2 & 0 (0) & 0 (0) & 0.21 (0.15) & 0.41 (0.28) & 98.57 (0.98) & 97.53 (1.66)  &  2.15 (0.05)  \\ 
		BIP & S2 & 0 (0) & 0 (0) & 0.66 (0.26) & 0.83 (0.36) & 95.71 (1.68) & 94.95 (2.1) & 2.20 (0.05) \\ 
		SELPCCA$_1$ & S2 & 22.00 (8.75) &16.33 (7.51) &.02 (.01)&.04 (.02)& 80.17 (7.81) &85.91 (6.37)&2.13 (0.04)  \\
		SELPCCA$_4$&S2&46.67	(4.04)&	44.67	(3.75)&	5.82	(0.42)&	7.31	(1.38)&	43.05	(3.15)	&41.83	(2.15)	&2.14	(0.04)\\		
		CCAReg$_1$ & S2 & 85.83 (1.87) &86.00 (1.79) &1.54 (.63) &1.27 (.47)& 19.52 (1.77) &19.76 (1.47) &2.03 (0.04)\\
		CCAReg$_4$&S2&89.00	(1.53)&	89.00	(1.30)&	4.34	(1.65)&	3.89	(1.50)	&12.37	(0.62)&	13.06	(0.73)&	2.10	(0.04)\\ 		
		FusedCCA$_1$&S2&0.00	(0.00)	&0.00	(0.00)&	17.14	(0.33)&	17.13	(0.39)	&42.79	(0.47)&	42.83	(0.53)&	2.21	(0.04)\\
		FusedCCA$_4$&S2&0.00	(0.00)&	0.00	(0.00)&	27.27	(1.22)&	24.51	(1.30)&	32.44	(0.97)&	34.94	(1.06)&	2.25	(0.04)\\		
		\hline
		BIPnet & S3 & 0 (0) & 0 (0) & 1.45 (0.15) & 1.45 (0.13) & 94.98 (0.48) & 94.98 (0.44)& 2.08 (0.04)  \\ 
		BIP & S3 & 0 (0) & 0 (0) & 0.62 (0.15) & 0.61 (0.12) & 97.81 (0.52) & 97.83 (0.43)&  2.10 (0.05) \\ 
		SELPCCA$_1$ & S3 & 8.50 (5.86) &8.42 (5.81) &0.16 (.06) &.14 (.05)& 92.02 (5.07) &92.20 (5.04)&2.05 (0.04) \\
		SELPCCA$_4$&S3&39.50	(2.95)&	29.00	(2.09)&	5.63	(0.28)&	5.64	(0.45)&	59.55	(2.12)&	66.94	(1.65)&	2.04	(0.04)\\ 		 		
		CCAReg$_1$ & S3 & 83.00 (2.50) &82.92 (2.60) &2.42 (.68)&2.51 (.74)& 23.21 (2.71) &23.02 (2.87) &1.98 (0.04) \\
		CCAReg$_4$&S3&88.25	(1.97)&	89.58	(1.93)&	3.42	(1.30)&	3.47	(1.33)&	15.74	(1.80)&	13.80	(1.4)6&	2.02	(0.05)\\	
		FusedCCA$_1$&S3&0.00	(0.00)	&0.00	(0.00)&	12.70	(3.01)&	13.89	(3.23)&	71.51	(2.31)&	69.80	(2.43)&	2.18	(0.04)\\
		FusedCCA$_4$&S3&0.00	(0.00)&	0.00	0.00&	25.47	2.31&	29.16	3.59&	53.62	2.34&	51.52	2.74&	2.22	(0.05)\\ 		
		\hline
		BIPnet & S4 & 0 (0) & 0 (0) & 2.61 (0.42) & 2.62 (0.4) & 77.37 (2.96) & 77.09 (2.81)& 2.24 (0.05) \\ 
		BIP & S4 & 0 (0) & 0 (0) & 2.58 (0.42) & 3.14 (0.45) & 77.68 (3.04) & 74.04 (3.24)& 2.27 (0.04) \\ 
		SELPCCA$_1$ & S4 & 51.32 (9.50) &42.89 (9.42) &0.10 (0.10)&0.00 (0.0)& 54.36 (8.08) &63.42 (8.15)&2.16 (0.04)  \\
		SELPCCA$_4$&S4&36.58	(4.97)&	29.47	(4.81)&	8.80	(2.96)	&5.67	(0.50)	&39.84	(2.88)	&45.22	(3.24)	&2.13	(0.05)\\ 		
		CCAReg$_1$ & S4 & 77.63	(1.87)&	78.68	(1.85)&	3.12	(0.68)	&2.71	(0.67)	&22.80	(1.15)	&23.49	(1.43)	&2.01	(0.05)\\
		CCAReg$_4$&S4&84.74	(2.12)&	83.42	(2.36)&	7.09	(2.34)&	6.66	(2.28)&	13.05	(1.34)&	13.68	(1.28)&	2.09	(0.06) \\		
		FusedCCA$_1$&S4&0.00	(0.00)&	0.00	(0.00)&	19.51	(2.22)&	19.64	(2.53)&	30.39	(1.03)&	30.55	(1.07)&		2.58	(0.06)\\
		FusedCCA$_4$&S4&0.00	(0.00)&	0.00	(0.00)&	28.76	(2.99)&	30.99	(3.34)&	23.64	(1.40)&	22.43	(1.39)&	2.52	(0.05)\\ 		
		\hline
		BIPnet & S5 & 54.8 (3.64) & 50.5 (3.45) & 0.19 (0.05) & 0.25 (0.06) & 60.3 (3.34) & 64.42 (2.99)& 2.35 (0.21) \\ 
		BIP & S5 & 55.65 (3) & 52.65 (2.96) & 0.24 (0.08) & 0.32 (0.1) & 59.91 (2.77) & 62.63 (2.77)&  2.64 (0.22) \\ 
		SELPCCA$_1$ & S5 & 93.40(1.55)&	93.30(1.85)&0.06(0.06)&	0.00	(0.00)&	11.60	(2.52)&	11.57	(3.00)&		2.72	(0.13)\\
		SELPCCA$_4$&S5&53.10	(6.56)&	53.75	(4.26)	&8.16	(3.02)&	1.84	(1.06)&	48.80	(3.35)&	58.35	(3.17)&	1.81	(0.04)\\		
		CCAReg$_1$ & S5 & 88.85 (1.81) & 87.95 (1.89) &1.44 (0.52)&1.40(0.50)&  19.83 (2.30) &21.51 (2.44)&2.26 (.14)  \\
		CCAReg$_4$&S5&78.90	(2.79)&	77.65	(2.38)&	2.99	(1.26)&	2.86	(1.29)&	30.01	(1.91)&	32.29	(1.66)&	1.63	(0.04)\\
		FusedCCA$_1$&S5&0.00	(0.00)&	0.00	(0.00)&	3.47	(0.81)	&2.31	(0.55)	&93.89	(1.37)&	95.77	(0.97)&		2.72	(0.11)\\
		FusedCCA$_4$&S5&0.00	(0.00)&	0.00	(0.00)&	7.72	(1.12)&	6.31	(1.02)&	87.21	(1.62)&	89.30	(1.48)&	2.01	(0.05) \\		
		\hline
		\hline
	\end{longtable}	
\end{scriptsize}

\subsection{Sensitivity results}
We assess the impact of the choice of hyperparameters $q_{\nu}$ on the posterior inference as follows. We applied our BIPnet method at a range of values to data generated from settings 1 and 5 described in the main manuscript. We repeated these analysis 20 times (Tables \ref{sensi1} and \ref{sensi1}). We observed negligible variations in our results across all scenarios. However, in setting 5 in particular, higher values of $q_{\nu}$ increased the number of false positive and decreased the number of false negative. From tables \ref{sensi3} and \ref{sensi4}, when $r$ is larger, while the MSE is worst, the variable selection performance is getting better in setting 5.
\begin{table}[H]
	\centering
	\caption{BIPnet: Sensitivity results for the hyperparameter $q$, the prior probability for variable selection. S1 stands for setting 1, S5 stands for setting 5,  FNR1 stands for  false negative rate for $\bm{X}^1$, FPR1; false positive rate for $\bm{X}^1$.  F11 is F-measure for $\bm{X}^1$. AUC1 is the area under the ROC curve to assess variable selection for $\bm{X}^1$. 
		\label{sensi1}
	}
	\scriptsize
	\begin{tabular}{ll|llll}
		\hline
		Setting & $q_{\nu}$ & FNR1 & FPR1 & F11 & AUC1 \\ 
		\hline
		S1 & 0.025 & 0 (0) & 0.03 (0.02) & 99.94 (0.04) & 1 (0) \\ 
		S1 & 0.05 & 0 (0) & 0.08 (0.05) & 99.84 (0.09) & 1 (0) \\ 
		S1 & 0.1 & 0 (0) & 0.21 (0.09) & 99.59 (0.18) & 1 (0) \\ 
		S1 & 0.2 & 0 (0) & 4.1 (1.36) & 93.41 (2.09) & 1 (0) \\ \hline
		S5 & 0.025 & 61.17 (3.5) & 0.01 (0.01) & 54.16 (3.72) & 0.88 (0.02) \\ 
		S5 & 0.05 & 65.85 (3.03) & 0.07 (0.03) & 49.34 (3.45) & 0.82 (0.02) \\ 
		S5 & 0.1 & 43.39 (3.54) & 0.25 (0.08) & 70.48 (3.07) & 0.84 (0.01) \\ 
		S5 & 0.2 & 32.81 (3.73) & 8.55 (1.78) & 66.01 (2.02) & 0.81 (0.02) \\ 
		\hline
	\end{tabular}
\end{table}

\begin{table}[H]
	\centering
	\scriptsize
	\caption{BIPnet: Sensitivity results for the hyperparameter $q$, the prior probability for variable selection. S1 stands for setting 1, S5 stands for setting 5,  FNR2 stands for  false negative rate for $\bm{X}^2$, FPR2; false positive rate for $\bm{X}^2$.  F12 is F-measure for $\bm{X}^2$. AUC2 is the area under the ROC curve to assess variable selection for $\bm{X}^2$. MSE is mean square error.
		\label{sensi2}
	}
	\begin{tabular}{ll|lllll}
		\hline
		Setting & $q_{\nu}$ & FNR2 & FPR2 & F12 & AUC2 & MSE \\ 
		\hline
		S1 & 0.025 & 0 (0) & 0.03 (0.03) & 99.94 (0.05) & 1 (0) & 1.92 (0.04) \\ 
		S1 & 0.05 & 0 (0) & 0.04 (0.02) & 99.92 (0.04) & 1 (0) & 1.91 (0.03) \\ 
		S1 & 0.1 & 0 (0) & 0.24 (0.1) & 99.54 (0.2) & 1 (0) & 2.43 (0.34) \\ 
		S1 & 0.2 & 0 (0) & 0.96 (0.28) & 98.17 (0.51) & 1 (0) & 1.92 (0.03) \\ \hline
		S5 & 0.025 & 68.25 (2.86) & 0.08 (0.07) & 46.85 (3.19) & 0.88 (0.02) & 3.72 (0.23) \\ 
		S5 & 0.05 & 64.09 (2.5) & 0.05 (0.03) & 51.83 (2.64) & 0.83 (0.02) & 3.44 (0.17) \\ 
		S5 & 0.1 & 46.55 (3.37) & 0.3 (0.09) & 67.84 (2.92) & 0.85 (0.01) & 3.07 (0.11) \\ 
		S5 & 0.2 & 30.18 (3.19) & 9.4 (1.97) & 67.22 (2.51) & 0.83 (0.02) & 3.02 (0.11) \\ 
		\hline
	\end{tabular}
\end{table}

\begin{table}[H]
	\centering
	\scriptsize
	\caption{BIPnet: Sensitivity results for  $r$, the number of components. S1 stands for setting 1, S5 stands for setting 5,  FNR1 stands for  false negative rate for $\bm{X}^1$, FPR1; false positive rate for $\bm{X}^1$.  F11 is F-measure for $\bm{X}^1$. AUC1 is the area under the ROC curve to assess variable selection for $\bm{X}^1$. 
		\label{sensi3}
	}
	\begin{tabular}{ll|llll}
		\hline
		Setting & $r$ & FNR1 & FPR1 & F11 & AUC1 \\ 
		\hline
		S1 & 3 & 0 (0) & 0.03 (0.02) & 99.94 (0.04) & 1 (0) \\ 
		S1 & 4 & 0 (0) & 0.08 (0.05) & 99.84 (0.09) & 1 (0) \\ 
		S1 & 6 & 0 (0) & 0.08 (0.05) & 99.84 (0.09) & 1 (0) \\ 
		S1 & 8 & 0 (0) & 0.18 (0.07) & 99.64 (0.14) & 1 (0) \\ \hline
		S5 & 3 & 70.94 (1.54) & 0.01 (0.01) & 44.62 (1.78) & 0.82 (0.02) \\ 
		S5 & 4 & 65.85 (3.03) & 0.07 (0.03) & 49.34 (3.45) & 0.82 (0.02) \\ 
		S5 & 6 & 63.63 (2.47) & 0.12 (0.07) & 52.28 (2.61) & 0.93 (0.01) \\ 
		S5 & 8 & 50.99 (4.9) & 0.37 (0.16) & 62.49 (3.96) & 0.96 (0.01) \\ 
		\hline
	\end{tabular}
\end{table}

\begin{table}[H]
	\centering
	\scriptsize
	\caption{BIPnet: Sensitivity results for  $r$, the number of components. S1 stands for setting 1, S5 stands for setting 5,  FNR2 stands for  false negative rate for $\bm{X}^2$, FPR2; false positive rate for $\bm{X}^2$.  F12 is F-measure for $\bm{X}^2$. AUC2 is the area under the ROC curve to assess variable selection for $\bm{X}^2$.  MSE is mean square error.
		\label{sensi4}
	}
	\begin{tabular}{ll|lllll}
		\hline
		Setting & $r$ & FNR2 & FPR2 & F12 & AUC2 & MSE \\ 
		\hline
		S1 & 3 & 0 (0) & 0.01 (0.01) & 99.97 (0.03) & 1 (0) & 1.95 (0.06) \\ 
		S1 & 4 & 0 (0) & 0.04 (0.02) & 99.92 (0.04) & 1 (0) & 1.91 (0.03) \\ 
		S1 & 6 & 0 (0) & 0.18 (0.08) & 99.65 (0.16) & 1 (0) & 2.2 (0.2) \\ 
		S1 & 8 & 0 (0) & 0.14 (0.07) & 99.73 (0.13) & 1 (0) & 1.96 (0.04) \\ \hline
		S5 & 3 & 69.24 (2.41) & 0.03 (0.02) & 46.07 (2.69) & 0.81 (0.02) & 3.81 (0.28) \\ 
		S5 & 4 & 64.09 (2.5) & 0.05 (0.03) & 51.83 (2.64) & 0.83 (0.02) & 3.44 (0.17) \\ 
		S5 & 6 & 50.82 (5.16) & 0.21 (0.1) & 62.5 (4.55) & 0.95 (0.01) & 6.14 (1.72) \\ 
		S5 & 8 & 42.4 (5.27) & 0.34 (0.12) & 69.66 (4.33) & 0.95 (0.01) & 5.37 (1.99) \\ 
		\hline
	\end{tabular}
\end{table}
\newpage
\section{Grouping information of genes and SNPs}\label{secgroups}
{
	\scriptsize
	\begin{longtable}{r|l|}
		\caption{List of genes within each IPA gene network. Gene names in bold face have the highest averaged MPPs ($>$0.85) across the 20 random splits.}\label{tableC1}\\
		\hline
		\multicolumn{1}{c|}{\textbf{Network \#}} & \multicolumn{1}{c|}{\textbf{Gene lists}}\\\hline
		\endfirsthead
		\endhead
		\hline \multicolumn{2}{l|}{{Continued on next page}} \\ 
		\endfoot
		\hline
		\endlastfoot
		1 & BSG;CPM;FBXO7;$\bf{KCTD12}$;LGALS3BP;LRRC37B;MPG;MRPL18;$\bf{MRPL22}$;$\bf{NAMPT}$;NEK4;NKD2;NOL7\\
		&PEG10;RAB7A;RAD18;RBM39;RPL35;RPS23;RPS29;$\bf{RPS3}$;$\bf{RPS4X}$;TMED3;TRIM8;UBE2A;VHL;XKR8  \\ 
		2 &ALDH7A1;CARD6;CRTAP;$\bf{CSTB}$;DTX3;EDA;EDARADD;$\bf{EMP3}$;FBXL16;GIMAP5;GPM6A;$\bf{GRN}$;MEFV;\\
		& PCDHAC2;PI3;$\bf{RAB10}$;RAB11FIP5;RAB37;RAB3C;RAB8A;RFTN1;SEC11C;$\bf{SLPI}$;SPEF2;SPG11;ST3GAL5\\
		&TMOD2;$\bf{TPMT}$;ZNF274 \\ 
		3 & ACP2;ATF7;BCLAF1;CCND1;CMAS;CTSE;CTSH;DAZ2;DCTD;DDX46;EEF1B2;EIF5A2;$\bf{GPBAR1}$\\
		&HACE1;HDX;KIF4A;LUC7L;LUC7L2;NCAPD3;NIF3L1;TBC1D7;TRPM8;ULK2;UTP6;ZNF181 \\ 
		4 & $\bf{ABI3}$;APIP;ATXN7L2;CCDC82;CHGA;CHGB;CLPTM1L;DSE;$\bf{ENY2}$;$\bf{GNG10}$;GNG4;GRK4;HOMER2;\\
		&HOMER3;$\bf{ITPR3}$;$\bf{MAK}$;MAPK1;NEIL3;NEK2;NR2C1;PATL1;RGS8;RSBN1;RSBN1L;SCG3;SPTBN4\\ 
		5 & ANGPTL3;BCHE;BRMS1L;CBX8;CHST12;CHST5;DPEP2;FBXO2;FUZ;GCKR;HS6ST3;$\bf{LY86}$;$\bf{LY96}$\\
		&MARVELD3;MCOLN3;MID2;NOVA1;PNOC;PON2;RBP1;$\bf{SLC12A9}$;SULT1C3;$\bf{TNFAIP8L1}$;ZNF526;ZNF785  \\ 
		6 & ANKRD34A;ATP13A1;CCDC136;DCLK1;FARP2;GOLM1;LHB;LRRTM1;MAP3K10;NCK1;NUFIP1;NUP62CL\\
		&PIGW;$\bf{PKN2}$;PNPLA5;PSMF1;RABL2A;$\bf{RPL14}$;RPL24;RPL37;SLC13A2;SLC39A4;SNURF;TWISTNB\\
		&$\bf{UBA52}$;VPS13A;ZNRF4 \\ 
		7 & ALG1;ATP11B;DNASE1L1;DNMT3L;$\bf{FBXO6}$;GGH;HSPD1;KLF12;MAX;MCM5;MDH1;PASK;PDHA2\\
		&$\bf{PPM1B}$;RPN2;SCRIB;SDK2;SYNE2;TMEM201;TOMM34;TTC13;VCAM1;$\bf{YY1}$ \\ 
		8 & ATM;BCL11A;CCNC;CPEB4;CSPP1;DCLRE1C;DHX40;FAM91A1;KLHL26;MED11;MED23;MED31;\\
		&MED4;MTHFR;MYH13;NIN;NRBP1;PERP;RASAL1;RIF1;SERPINB9;SYF2;TGIF1;TNRC6B;ZNF592 \\ 
		9 & ACTL6A;BCL7A;CCDC71;CPSF6;$\bf{CPVL}$;;DDX19A;EDF1;FLII;HIST1H4K;HMGN1;LSAMP;MGMT;MIER2;\\
		&MIF4GD;NCOA3;NEGR1;PARP15;PCGF1;PCGF2;PRAME;SFMBT1;SMARCAD1;SPTY2D1;TAF1;UHRF1 \\ 
		10 & $\bf{AKR1A1}$;BIRC7;BTBD3;CAB39L;CD160;COX4I2;$\bf{COX5A}$;$\bf{COX6B1}$;$\bf{COX7A2}$;COX8C;$\bf{CYP1B1}$\\
		&GSTT1;HSPB6;HTATIP2;KIF15;$\bf{MCL1}$;$\bf{MGST2}$;$\bf{NDUFAF2}$;$\bf{NDUFB2}$;OPTC;$\bf{SLC15A4}$\\
		&SNRK;STARD9;$\bf{UQCRQ}$ \\ 
		11 & ADAM19;ADAM2;$\bf{ALDOC}$;$\bf{ASPHD2}$;BCAN;BTC;CDH2;CNFN;COL19A1;DCN;EN2;ITGB1;METRNL;MME\\
		&MMP10;MMP12;MYOCD;$\bf{PSAP}$;PYGB;$\bf{PYGL}$;RNASE4;ST6GAL1;THBS1;$\bf{VCAN}$ \\ 
		12 & $\bf{ARHGAP4}$;C1QTNF2;$\bf{CHN2}$;DEFB110;DEFB114;DEFB126;INSL4;ITLN1;NDE1;NLRC5;OTUD1;\\
		&$\bf{PCSK1N}$;PRR7;$\bf{PSMA4}$;$\bf{PSMB10}$;PSMB8;PSMD5;PSMD6;PSMD9;PSME1;RAD23A;$\bf{TCF7}$;TSNARE1 \\ 
		13 & ABCA8;CHD5;CLGN;CLN5;DGKA;$\bf{FAM133B}$;GAA;HEMK1;$\bf{HIVEP1}$;LIPH;MEF2A;MTRF1L\\
		&$\bf{MZF1}$;NAGLU;NCKAP1;NPHP4;SRGAP1;SRGAP3;TIAM2;TNFRSF17;TRMT1;WASF3;ZMAT3 \\ 
		14 & ARL4D;CDH13;COL13A1;CRABP2;CST6;CTNNB1;IDI1;LYPD6;$\bf{MAN2B1}$;MDN1;PCK1;\\
		&POLR3B;PRKCSH;QPCT;RSPO1;SEC31A;SP6;$\bf{SSH1}$;SYT12;TBC1D2;UNC13C;ZIC2 \\ 
		15 & AIP;ASGR2;CCDC86;$\bf{CYBA}$;DIDO1;DNAJC3;DNAJC6;DNAJC9;HABP4;MICB;MT1H;\\
		&NADSYN1;PCMT1;PLOD1;$\bf{SECISBP2}$;SEPHS2;SHKBP1;TJP2;TMOD1;TPM2  \\ 
		16 & ATG2B;B3GALT2;C1QTNF9;CD72;ELK4;IL11RA;IL17RD;$\bf{IL1RAP}$;IL1RL2;IL22RA1;\\
		&JPH4;MAP3K7;MRS2;MYO10;MYOM1;$\bf{NCF4}$;SOCS5;TACC1;TDRD7;$\bf{TGFBI}$ \\ 
		17 & CAPG;CAPZA3;COL17A1;COL4A3BP;DAG1;EGFLAM;EMILIN1;$\bf{EMILIN2}$;$\bf{ITGA6}$;ITGA7\\
		&KRT18;KRT3;KRT6A;KRT86;MAT1A;PRX;RER1;RFX4;SGCA;TNS1;UCP3;UTRN \\ 
		18 & ABAT;ADCY1;ANAPC13;BUB3;CAPN10;CDC26;EIF2B5;FOXE1;FOXK2;HADH;KLF7;\\
		&PGM1;RASSF6;SCAMP1;SLC25A27;WDR19 \\ 
		19 & ADRA1A;BDKRB2;CCKAR;CCKBR;CCR9;GABBR1;GJC2;GPR157;GPR34;\\
		&GPR35;HTR1F;RGS1;RGS11;VN1R1 \\ 
		20 & ADCY3;AMOTL2;ARL4C;$\bf{CLDN23}$;CSTF3;FIP1L1;FOXA1;GAL;NRG1;PCDHA5;PHF2;\\
		&PLCD1;PLEKHG5;PUS7;RAB2A;$\bf{RAB2B}$;RAP1GDS1;ROM1;SREBF1;TSC22D3;WDR33;YTHDF2 \\ 
		21 & AKR1D1;DERL1;EPHX1;ERBB2;FLT1;KCNA4;KCNAB1;KLF6;KLHL21;LGALS4;MATK;\\
		&MRPS21;POLB;RET;RRAS2;SEC61A1;$\bf{SEC61G}$;SH2D5;$\bf{SYK}$;TYRO3 \\ 
		22 & ARNTL2;BLOC1S2;$\bf{DENND3}$;DTNBP1;$\bf{DUSP23}$;ERMAP;FBXL15;IFI16;KEL;NUDT11;\\
		&PPFIA1;PPM1K;PPP1R8;PTPN22;RDM1;RORC;RTP4;SESN3;$\bf{SLC2A11}$;SNX20;STAT6 \\ 
		23 &AMY2A;APLN;$\bf{CD33}$;DIS3;DOC2A;EIF3B;KCNK3;LYL1;$\bf{MT2A}$;PHACTR3;\\
		&PIK3C3;PTGDS;$\bf{SH3GLB1}$;STX8;$\bf{STXBP3}$;SYPL1;$\bf{VAMP8}$;VPS18;$\bf{VPS41}$ \\ 
		24 & C6;CDK5RAP3;$\bf{CR1}$;ECM2;GFRA1;HPN;KLK1;MAPKAPK5;MS4A1;PAPSS2;\\
		&PRELP;PRSS21;SERPINA1;SOS2;SPRED3;STK40;TFPI;THOP1;ZIC3 \\ 
		25 & $\bf{ANG}$;$\bf{AP2S1}$;$\bf{CD14}$;CHI3L1;CNR1;CSF3;$\bf{CXCR4}$;DYM;$\bf{FPR2}$;GIMAP4;GNAI3;GNLY\\
		&HRASLS;IRAK3;$\bf{NECAP2}$;PLSCR1;REEP6;RTN2;TCN1;TLR6 \\ 
		\hline
	\end{longtable}
}
\begin{table}[H] 
	\centering
	\scriptsize
	\caption{Biological characteristics of IPA gene networks.}\label{tableC2}
	\begin{tabular}{rl}
		\hline
		Network \#& Top Diseases and Functions \\ 
		\hline
		1 & [Cancer, Protein Synthesis, RNA Damage and Repair] \\ 
		2 & [Cell-To-Cell Signaling and Interaction, Dermatological Diseases and Conditions,\\
		&Organismal Injury and Abnormalities] \\ 
		3 & [Cellular Movement, Hematological System Development and Function, Immune Cell Trafficking] \\ 
		4 & [Cell Cycle, Endocrine System Disorders, Organismal Injury and Abnormalities] \\ 
		5 & [Auditory and Vestibular System Development and Function, Cellular Development, \\
		&Cellular Growth and Proliferation] \\ 
		6 & [Infectious Diseases, Protein Synthesis, RNA Damage and Repair] \\ 
		7 & [Amino Acid Metabolism, Cardiovascular System Development and Function, Cellular Function and Maintenance] \\ 
		8 & [Cell-mediated Immune Response, Cellular Development, Cellular Function and Maintenance] \\ 
		9 & [Cell Cycle, Cellular Assembly and Organization, DNA Replication, Recombination, and Repair] \\ 
		10 & [Cancer, Cell-To-Cell Signaling and Interaction, Inflammatory Response] \\ 
		11 & [Carbohydrate Metabolism, Connective Tissue Development and Function, Tissue Development] \\ 
		12 & [Gene Expression, RNA Damage and Repair, RNA Post-Transcriptional Modification] \\ 
		13 & [Cell Morphology, Cellular Assembly and Organization, Cellular Function and Maintenance] \\ 
		14 & [Cell Cycle, Embryonic Development, Organismal Development] \\ 
		15 & [Connective Tissue Disorders, Dermatological Diseases and Conditions, Developmental Disorder] \\ 
		16 & [Cancer, Cell Signaling, Cell-To-Cell Signaling and Interaction] \\ 
		17 & [Developmental Disorder, Hereditary Disorder, Ophthalmic Disease] \\ 
		18 & [Connective Tissue Disorders, Developmental Disorder, Hereditary Disorder] \\ 
		19 & [Cell Signaling, Cell-To-Cell Signaling and Interaction, Molecular Transport] \\ 
		20 & [Behavior, Hereditary Disorder, Nervous System Development and Function] \\ 
		21 & [Cancer, Gastrointestinal Disease, Organismal Injury and Abnormalities] \\ 
		22 & [Cell Cycle, Cell-To-Cell Signaling and Interaction, Cellular Growth and Proliferation] \\ 
		23 & [Cellular Assembly and Organization, Infectious Diseases, Molecular Transport] \\ 
		24 & [Humoral Immune Response, Inflammatory Response, Neurological Disease] \\ 
		25 & [Cell-To-Cell Signaling and Interaction, Cellular Function and Maintenance, \\
		&Hematological System Development and Function] \\ 
		\hline
	\end{tabular}
\end{table}

{ \scriptsize
	\begin{longtable}{r|l|}
		\caption{List of SNPS with their corresponding genes (or groups). SNP names in bold face have the highest averaged MPPs ($>$0.85) across the 20 random splits. }\label{tableC3}\\
		\hline
		\multicolumn{1}{c|}{\textbf{Genes}} & \multicolumn{1}{c|}{\textbf{List of SNPs}}\\\hline
		\endfirsthead
		\endhead
		\hline \multicolumn{2}{l|}{{Continued on next page}} \\ 
		\endfoot
		\hline
		\endlastfoot
		HDAC9 & rs10486302;rs10486301;rs10236525;rs213273;rs213274;rs213276;rs3814991;rs2269749;\\
		&rs11764116;rs2249817;rs11768780;rs10248565;rs11984041;rs2074633;rs2023938;rs28688791 \\ 
		ANKS1A & rs2273007;rs6918872;rs12205331;rs1737727;rs847851;rs847848;rs7740148;\\
		&rs2140418;rs3800435;rs2689088;rs13210323;rs7742443;rs17609940;rs820065 \\ 
		SMG6 & rs9914578;rs4790881;rs11658881;rs17761734;rs2281727;rs216172;rs10852932;\\
		&rs216219;rs216182;rs1885987;rs216193;rs2224770;rs3760229 \\ 
		P4HA2 & rs3761661;rs162887;rs460089;rs460271 \\ 
		BCAP29 & rs10953541;rs3801944;rs10273733 \\ 
		FLT1 & rs7326277;rs2296188;rs7987291;rs3794400;rs1408245;rs17625898;rs7995976;rs3751397;\\
		&rs7983774;rs9319428;rs9513097;rs9508025;rs9508026;rs2387634;rs10507385;rs9551471;\\
		&rs585421;rs748253;rs600640;rs678714;rs9554312 \\ 
		CXCL12 & rs2839695;rs2839689;rs10793538;rs11592974;rs7092453;rs17540465;rs2146807;rs928565;\\
		&rs1147882;rs266109;rs1147879;rs266108;rs7918568 \\ 
		RASD1 & rs711352;rs11545787 \\ 
		RAI1 & rs4925102;rs7224617;rs8071107;rs11654482;rs11654526;rs4636969;rs9914733;rs9907986;\\
		&rs752579;rs3803763;rs8067439;rs4925112;rs35686634;rs3818717;rs11654081;rs4925114 \\ 
		PHACTR1 & rs7768030;rs9357503;rs9349350;rs9296486;rs9296488;rs9381439;rs9296494;rs17617584;\\
		&rs1332844;rs7750679;rs9369640;rs6906890;rs10485363;rs1223531;rs9357620;rs4715166 \\ 
		GGCX & rs2028898;rs10179904;rs2592551;rs699664;rs762684;rs6738645;rs17026452;rs12714145;rs7568458 \\ 
		VAMP8 & rs3755008;rs3731828;rs1058588;rs1010 \\ 
		PLG & rs9458011;rs3757019;rs9295131;rs4252109;rs14224;rs783147;rs2295368;rs13231;rs813641;\\
		&rs4252159;rs4252170;rs783182;rs783173;rs4252051;rs4252052;rs4252053;rs2314852;rs1950562 \\ 
		PEMT & rs16961845;rs4646396;rs750546;rs1531100;rs11658311;rs4646356;rs4244598;rs12325817;rs3760187;rs4646343;rs4646341 \\ 
		COL4A2 & rs4773143;rs4773144;rs7986871;rs3809346;rs4773161;rs12876517;rs7319311;rs9515203;rs1927343;\\
		&rs1927349;rs7334986;rs60212072;rs7983487;rs7984937;rs7984100;rs7983979;rs4771680;rs7489705;\\&rs4103;rs59905747;rs45612833;rs7326449;rs4132608;rs56676181;rs75082326;rs76425569;rs3803237;\\
		&rs3803236;rs3825490;rs72657953;rs1983931;rs1983932;rs1927350;rs3803232;rs9559814;rs9583500;\\
		&rs9515229;rs9559818;rs9301457;rs4773194;rs9521803;rs3803228;rs2296852;rs11839527;rs2296851;\\
		&rs413756;rs402661;rs403839;rs2296849;rs421177;rs2274544;rs2391833;rs9559826 \\ 
		EDNRA & rs1801708;rs10305860;rs6841473;rs908581;rs2048894;rs5333;rs5334;rs6842241;rs6841581 \\ 
		MRAS & rs1199338;rs1199337;rs4678260;rs2347252;rs1720819;rs2306374;rs3732837;rs9818870 \\ 
		UBE2Z & rs46522;rs15563;rs1057897 \\ 
		IL6R & rs12083537;rs1386821;rs6684439;rs7518199;rs7521458;rs4845622;rs4393147;rs4453032;rs4845623;\\
		&rs4537545;rs12730935;rs61812598;rs7529229;rs4845625;rs4845626;rs6689393;rs4129267;rs4845374;\\
		&rs2228145;rs11265618;rs10159236;rs10752641;rs4329505;rs6698040;rs4240872;\\
		&rs4509570;rs4341355;rs2229238;rs7514452;rs4379670 \\ 
		PLPP3 & rs17114036;rs11206831;rs17114046;rs1759752;rs1930760;rs10493211;rs10736393;rs12566304;rs914830 \\ 
		KCNK5 & rs10947789;rs2815118 \\ 
		SRR & rs408067;rs3744270 \\ 
		ZEB2 & rs12105918;rs13382811;rs6755392 \\ 
		PDGFD & rs10895547;rs2220377;rs12574463;rs1386751;rs1948122;rs488753;rs7950273 \\ 
		COL4A1 & rs681884;rs2275842;rs2275843;rs16975424;rs2298240;rs2298241;rs3742207;rs1213026;rs652572;rs1778817;rs589985;\\
		&rs2289800;rs2289799;rs1000989;rs874203;rs874204;rs1975514;rs1562173;rs2305081;rs503053;rs2131939;\\
		&rs10492497;rs16975491;rs16975492;rs7329411;rs72654112;rs2305080;rs61749897;rs9521638;rs2241967;\\
		&rs683309;rs496916;rs668847;rs668861;rs648263;rs7327728;rs648705;rs648735;rs482757;rs677877;\\&rs9588116;rs598893;rs7333008;rs7332120;rs7333204;rs645114;rs2166207;rs9521649;rs2166208;\\
		&rs76574181;rs9521650;rs41275104;rs41275106;rs12428227;rs494558;rs7987982 \\
		SLC22A4 & $\bf{rs35260072}$;rs3792876;rs3805665;rs3792880;rs3805668;rs13179900;rs270608;rs270607\\
		&rs2073838;rs2073839;rs3828673;rs3792885;$\bf{rs272842}$;rs3761659;rs3805673;rs273915;$\bf{rs272893}$;$\bf{rs272889}$\\
		&rs272887;rs273909;rs272882;$\bf{rs272879}$;rs272873;rs272872;rs2306772;$\bf{rs1050152}$  \\ 
		SLC22A5 & $\bf{rs2631367}$;rs13180043;rs13180295;rs3788987;rs2631362;rs2631359;rs4646301;rs274571;rs2073642;rs183898\\
		&$\bf{rs17622208}$;rs4646305;$\bf{rs274559}$;$\bf{rs274558}$;$\bf{rs274557}$;$\bf{rs2073643}$;rs274554;rs274553;rs274551;rs274549;rs274547 \\  
		AL008729.1 & rs202072;rs3817737;rs578290 \\ 
		COL4A2-AS2 & rs9515217;rs9515218;rs9555703;rs9515219;rs9521781;rs9521782;rs7990214;\\
		&rs7990383;rs4773186;rs41275110;rs7992330 \\ 
		TEX41 & rs900554;rs11681525;rs786255;rs2381683;rs1371048;rs10496971 \\ 
		MIR3936HG & rs2631372;rs4646298;rs2631369;rs2631368 \\ 
		\hline
	\end{longtable}
}

\section{Application to the ASCVD data}

This Section provides how we pre-process our data and some supplementary materials from the application of methods to the ASCVD data.
\subsection{Data preprocessing}\label{dataprocess}
There were 340 patients with gene expression and SNP data, as well as clincal covariates to calculate their ASCVD score. The proportion of females was 66.47\% and that of males was 33.53\%; their ages ranged from 40 to 78 with mean age 52.78 years. The following clinical covariates were added as a third data set: age, gender, BMI, systolic blood pressure, low-density lipoprotein (LDL), and triglycerides. The gene expression data consisted of 38,624 probes. Genes with variance and entropy expression values that were respectively less than the 90th and 50th percentile were removed. This resulted in 1,398 genes.

For the SNP data, we started with genes reported by  \cite{Kullo2016} that were found to be associated with ASCVD. We retrieved 1,152 SNPs in these gene regions from the NCBI dbSNP database but we had data on 718 SNPs, including 22 of the 28 SNPs reported in \cite{Kullo2016}.  It should be noted that the remaining SNPs  may or may not be associated with ASCVD. For SNP quality control, we included only SNPs that had call rate $=100\%$, were in Hardy-Weinberg equilibrium (P-value $\ge 1 \times 10^{-5}$), and had minor allele frequency (MAF) $\ge 1\%$. At the end of the filtering process, we had $429$ SNPs for the analyses. We assumed an additive genetic model in which the genotypes were coded ``0'' for non-risk allele homozygotes (alleles with most frequency in our data), ``1'' for heterozygotes, and ``2'' for risk-allele homozygotes (minor alleles). We treated the genetic data as continuous. The gene expression and SNP data were each standardized to have mean zero and standard deviation one for each feature. 

\begin{figure}[H]
	\caption{Marginal posterior probabilities of genes for each the of the 4 components  using method BIPnet+Cov for a random split.}\label{figure7}
	\centering
	\includegraphics[scale=0.60]{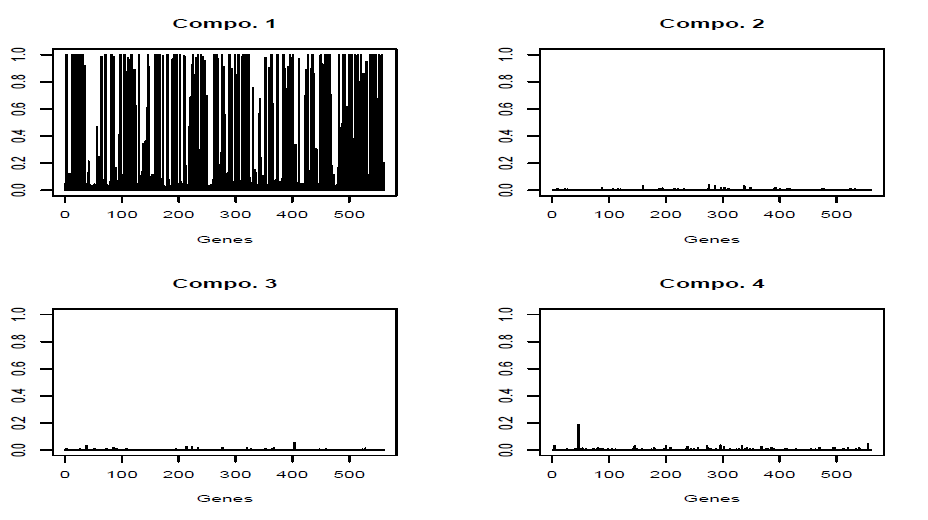}
\end{figure}

\begin{figure}[H]
	\centering
	\caption{Marginal posterior probabilities of SNPs for each the of the 4 components  using method BIPnet+Cov.}\label{figure8}
	\includegraphics[scale=0.60]{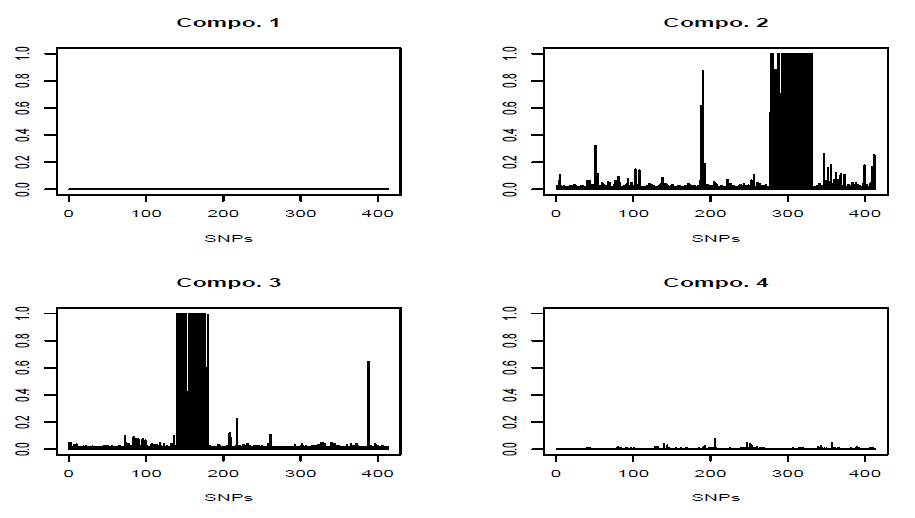}
\end{figure}
\begin{figure}[H]
	\centering
	\caption{Marginal posterior probabilities of  IPA gene networks (from component 1) using method BIPnet+Cov.}\label{figure9}
	\includegraphics[scale=0.65]{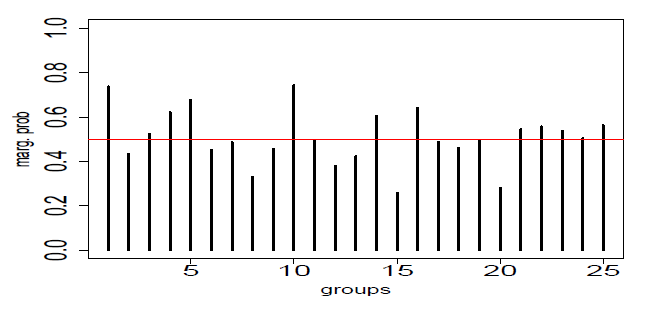}
\end{figure}

\begin{figure}[H]
	\centering
	\caption{Marginal posterior probabilities of SNP groups (i.e genes) using method BIPnet+Cov. The plots on the left and right are the MPPs from component 3 and 4 respectively.}\label{figure10}
	\includegraphics[scale=0.65]{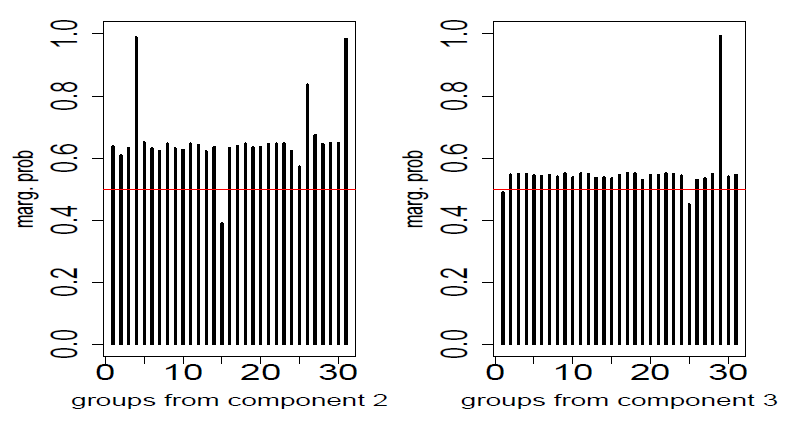}
\end{figure}


\end{document}